\documentclass{article}

\usepackage[style=numeric]{biblatex}
\usepackage{arxiv}
\bibliography{references} % Please set the right name for your bib file
\usepackage[utf8]{inputenc} % allow utf-8 input
\usepackage[T1]{fontenc}    % use 8-bit T1 fonts
\usepackage{hyperref}       % hyperlinks
\usepackage{url}            % simple URL typesetting
\usepackage{booktabs}       % professional-quality tables
\usepackage{amsfonts}       % blackboard math symbols
\usepackage{nicefrac}       % compact symbols for 1/2, etc.
\usepackage{microtype}      % microtypography
\usepackage{lipsum}
\usepackage{subfigure}

\usepackage{amsmath}
\usepackage{graphicx}
\usepackage{booktabs}
\graphicspath{{Images/}}
\usepackage{tikz}
\usepackage{quiver}
\usepackage{multicol}
\usepackage{multirow}
\usepackage{moreverb,url}
\usepackage[table]{xcolor}

\usepackage[T1]{fontenc}

\usepackage{newunicodechar}
\newunicodechar{−}{-}

\bibliography{references}
\title{\textbf{Normative feeling: socially patterned affective mechanisms }}

\author{Stavros Anagnou, Daniel Polani, Christoph Salge \\
   Adaptive Systems Research Group, Department of Computer Science, University of Hertfordshire\\
   \textit{s.anagnou@herts.ac.uk}  | \textit{d.polani@herts.ac.uk} | \textit{c.salge@herts.ac.uk} 
}

\begin{document}

\maketitle
\begin{abstract} \normalsize
Breaking a norm elicits both material and emotional consequences, yet how this coupling arose evolutionarily remains unclear. We investigate this question in light of emerging work suggesting that normativity's building blocks emerged earlier in evolution than previously considered, arguing that normative processes should inform accounts of how even ancient capacities such as mood evolved. Using a definition of normative processes we developed, we created an agent-based model with evolvable affect in a shared resource dilemma, comparing competition (non-normative) versus punishment (normative) conditions. Critically, different mood mechanisms emerge under each condition. Under competition, agents evolve a "bad mood → consume more" response, creating a tragedy of the commons leading to resource depletion and population collapse. Under punishment, agents evolve a "bad mood → consume less" mechanism, where negative affect functions as an implicit signal of social sanction, promoting resource conservation. Importantly, once normative logic is imprinted through punishment, it creates an evolutionary pathway for mood-based signalling that operates without costly physical enforcement. Our findings demonstrate how normative processes enable social preferences to emerge in a distributed manner within psychological mechanisms, showing how normative processes reprogram cognitive and physiological systems by embedding cultural patterns into psychological dispositions.

\end{abstract}

% keywords can be removed
\keywords{Norm emergence, Affect, Emotion modelling, Distributed mechanism, Normative regularity, Social preferences, Norm definition, evolution of mood}

\vfill
\section{Introduction} 
Social norms are an incredibly important aspect of organising a group, as they allow for groups of individuals to coordinate and cooperate with one another. Example norms include, standing a certain distance away in conversation: Too close can be seen as invasive; too far might feel cold or regulating individual greed when managing a shared resource sustainably: too much is seen as selfish but too little is seen as too generous.
It has been shown that processes such as punishment and incentives are the key mechanisms underpinning the emergence of norms. They align incentives so that individuals follow norms as to avoid punishment or to gain some reward \cite{andrighetto_punish_2013,raihani_punishment_2012}. 
However, we know that this isn't the only reason individuals follow norms; individuals seem to become compelled to follow norms instinctually due to their feelings, even when explicit material punishment is not incurred \cite{frank_passions_1988}. For example, psychologist Stanley Milgram instructed his students to ask to take someone's seat on the subway to examine how people would react to the breaking of a norm. His students' attempts were largely unsuccessful, most citing they could not bring themselves to it. Stanley, therefore, decided to try for himself, but when he approached a seated passenger, he found himself stuck. "The words seemed lodged in my trachea and would simply not emerge", he said. Retreating, he berated himself: 'What kind of craven coward are you?" A few unsuccessful tries later, he managed to blurt out a request. "Taking the man's seat, I was overwhelmed by the need to behave in a way that would justify my request," he said. "My head sank between my knees, and I could feel my face blanching. I was not role-playing. I actually felt as if I were going to perish." This visceral account of norm breaking suggest that the norms out there in the world and our feelings are interwoven \cite{blass_man_2009}. This shows that the concern of how others will perceive us is deeply entrenched in our emotions.  The question we want to ask in this article is how this tight coupling between emotions and norms \cite{kao_computational_2023,rai_social_2024}, henceforth social preference, come about evolutionarily?

\subsection{Evolutionary perspective on norms, punishment and mood}

To address the emergence of social preferences in mood, we need to first consider how norms and their enforcement came about evolutionarily.

Since social norms feature mainly in human societies, many scholars have proposed that these norms arise from cognitive capacities unique to humans. For instance, some theorists argue that social norms are rooted in distinctively human abilities e.g. shared intentionality \cite{tomasello_moral_2020}, or a specialised human "norm psychology," \cite{henrich_secret_2016} shaped by a long history of gene–culture coevolution specific to hominins. Andrighetto et al. argue for a cognitive machinery to detect and reason upon norms that is characterised by a salience mechanism devoted to tracking how much a norm is prominent within a group \cite{andrighetto_punish_2013}. More recently, \cite{birch_toolmaking_2021} has suggested that social norms may have emerged from cognitive control systems initially selected for the manufacture of complex tools in early hominins \cite{birch_toolmaking_2021}.

However, these definitions of norms assume that human cognition is essential to norms, meaning we may overlook similar norm-like processes occurring in non-human animals or artificial systems through different cognitive means.

Therefore, others have proposed a more permissive definition that is psychologically agnostic, and therefore allows researchers to include all the different ways animals (human and non-human) may produce norms or “normative regularities”. For example, the normative regularities view \cite{westra_search_2024} or the axiomatic normativity view, which emphasise social processes that result in multiple behavioural equilibria \cite{anagnou_effect_2023}.

By taking a more inclusive and comparative approach, we may be better able to study the fundamental building blocks of normative cognition. This ultimately will allow us to better understand their evolutionary emergence, rather than assuming they came about in humans fully formed \cite{de_waal_towards_2010}. 

Therefore,  we see that (minimally) normative processes, such as punishment, occur not only in higher-level organisms like humans but in organisms across phyla \cite{raihani_punishment_2012}, i.e. simple organisms like insects \cite{wenseleers_enforced_2006} and bacteria \cite{huang_toxin-mediated_2023}. Bacillus velezensis SQR9 have bacterial free riders who benefit from the collective biofilm but do not contribute to it. They are punished by toxins released by the producers of the biofilm  \cite{huang_toxin-mediated_2023}. This also occurs in other species through punishment of cheating individuals by cooperator-produced antibiotics \cite{wechsler_understanding_2019, wang_quorum_2015}. The fact that minimally normative processes such as punishment occur across phyla means:

\begin{enumerate}
    \item We must not only consider normative punishment's effect on culture but also on the biological fitness of the organisms, e.g. for example, free rider phenotypes are selected out genetically, but the process is still fundamentally the same as it shapes behaviour, whether through genetics or culture.
    \item Given normative punishment's presence across different phyla, it must have emerged very early in evolution or evolved many times.
    \item This being the case, we must consider normative punishment in the evolution of simple and fundamental traits that arose early in evolution, e.g. affect (mood, emotions, hormonal processes),

\end{enumerate}

 We will focus on mood in this paper. We choose mood because many existing computational accounts of mood are individualistic or otherwise do not take into account normative processes, only simpler social but non-normative social processes like predation and competition  \cite{eldar_mood_2016, taborsky_towards_2021, trimmer_evolution_2013}. We further choose mood, also widely conserved across phyla, because mood-like mechanisms likely evolved very early in animal evolution, well before the emergence of humans or complex mammals \cite{bateson_agitated_2011,procenko_physically_2024}. It is one of the fundamental mechanisms of how animals navigate the world, since it gives a notion of what is “good” and “bad” for the organism \cite{jonas_critique_1953,simon_motivational_1967}.

Therefore, using our agent-based model with an evolvable mood mechanism, we investigate the effect of punishment on the kind of mood mechanism that evolves. 

In the process of doing this, we discover a candidate mechanism (mood in the presence of punishment) that results in the emergence of social preferences from individual-level selection alone. That is, how a mood mechanism itself can become altruistic via being patterned by normative punishment. Further, we also show how punishment allows individuals to align their mood mechanisms in a way so that they can affect each other's behaviour through affective signals and not physical sanctions, therefore avoiding the costly pitfalls of physical sanctions.

\section{Background}

\subsection{Evolution of social preferences}

Our attempt to model the normative influence on the evolution/patterning of affect is closely related to several other lines of research. For example, with regard to the emergence of social preferences, Frank \cite{frank_passions_1988} proposes that in social scenarios such as the Prisoner's Dilemma, some emotions described as "moral sentiments" commit a person to act contrary to their immediate self-interest \cite{frank_passions_1988}. For example, the predisposition to feel guilt commits a person to be altruistic, even if cheating were in their material interest. A person with the predisposition to get outraged after having been cheated is committed to punish the cheater, even if it is costly in material terms. Thereby, emotions such as guilt and anger act as "commitment devices" that alter the material incentives. Even so, there should be a material gain from having these emotions, otherwise they would not have evolved in the first place.

Computational studies have investigated how social preferences affect behaviour. Namely, by incorporating social preferences by hard-coding them into the agent model to see their effect on adaptive group behaviour e.g normative concern motivation \cite{alexander_staller_introducing_2001}, equity motivation \cite{de_jong_artificial_2008}. Although these studies illuminate how social preferences may affect behaviour in multi-agent settings, they do not explain the emergence of social preferences from purely material concerns, i.e. how they come about through individual-level selection. Other studies exist that do address the preconditions of internalising social preferences but assume trivial mappings binary between genome and behaviour, not based on biological mechanisms \cite{bowles_cooperative_2013,gavrilets_collective_2017}. For example, \cite*{cimpeanu_evolutionary_2025} study the material incentives that may lead to the evolution of guilt but uses a mechanism of guilt that is already functionally pre-specified and linked trivially to a binary genetic character instead of letting the functional relations of the emotion evolve by itself. Another study shows how preferences may be internalised in a psychologically realistic mechanism, but this finding relies on high-level cognitive capacities found only in humans \cite{andrighetto_norm_2010}.

In contrast to these other examples, we are interested in how social preference "emerges" and is encoded in our \textit{evolvable} model of mood out of purely material concern. That is, how individual-level selection, in a social setting, patterns a biologically plausible mechanism. 
In a similar approach to how other research looks at how social dynamics influence the evolution of affect e.g. how predator-prey interactions affect the neuroevolution of affect \cite{godin-dubois_spontaneous_2021,hesp_evolutionary_2021}. However, instead of prey-predator interactions, we want to clarify the effect of normativity, in the form of punishment, on the evolution of mood.

\subsection{Axiomatic Normativity: refining the large space of behavioural regularities}
%%shoudl i go through in detail for each *Jong et al, 2008 modify their agents utility function so that it engages in inequity avoidance, yet this is coded in instead of evolved instead of emerging from purely material concerns.*

In this section, we will explain minimally normative approaches, normative regularities \cite{westra_pluralistic_2022} and axiomatic normativity \cite{anagnou_effect_2023}.
“Social norms” refer to the rules of a group of people that mark out what is appropriate, allowed, required, or forbidden for various members in different situations. They are typically manifest in common behavioural regularities that are kept in place by social sanctions and social pressure \cite{kelly_psychology_2021}.

\cite{westra_pluralistic_2022} define “Normative regularities”  as “socially maintained patterns of behavioural conformity within a community”. Here, "patterns of behavioural conformity" correspond to readily discernible empirical behavioural
regularities that support robust predictions and generalisations and whose existence does not depend upon us knowing the underlying mechanisms that produce them. "Social maintenance" here is some form of positive reinforcement of the regularity or negative punishment for breaking away from the regularity. This criterion is needed since not all sources of behavioural conformity are normative. For example, if a group of animals all congregate at the same water source in an otherwise dry landscape, this need not imply that drinking at that water source is a social norm \cite{westra_search_2024}.
Therefore, there must be a process of “social maintenance” that maintains the behaviour so it is not purely a consequence of the environment (or as the authors call it environmental scaffolding).

Although this wider minimally cognitive definition is useful in terms of capturing norm-like behaviour in non-human animals, it is also very general and does not distinguish between important aspects of normative behaviour. Therefore, additional criteria are needed to distinguish between normative and behavioural regularities \cite{anagnou_effect_2023}. 

There could be social maintenance processes that incentivise individuals in a group to tend to one equilibrium, just as in the case with environmental scaffolding, and others where behaviours tend to many equilibria. For example, in competition, all individuals compete for a resource and are reacting to each other's behaviour; If an individual exploits a resource more than you, you may have to also exploit it more before they do. One can argue that this is social maintenance because individuals are implicitly punishing each other for not being as competitive as one another.
While competing, they are forced to increase their efforts more and more in order to outcompete others, thus the population will eventually arrive at a maximum possible value of behaviour, e.g. peak resource exploitation. Therefore, the outcome would be the same in all populations; only the greediest survive. 
This misses out on a key aspect of normativity in our view, which is a degree of arbitrariness \cite{anagnou_effect_2023}. We call processes like competition, social scaffolding, i.e. social process which pushes individuals towards the same outcome in every scenario. Another example is prey-predator arms races, where prey will get faster, which in turn forces the predator to be faster, but will ultimately tend to one equilibrium - the fastest possible speed.

Therefore, we introduce a multi-equilibria criterion which can distinguish between those two types of normative regularities. In general, we suggest the axiomatic norm definition, an approach where norms are defined by a set of criteria, which is as follows: 

\begin{enumerate}
    \item A behavioural regularity in a community of individuals.
    \item The behavioural regularity that is socially maintained (ala Westra et al, 2024)
    \item The behavioural regularity be one of multiple equilibria even when environmental conditions are kept constant. 
\end{enumerate}

This norm definition provides a lens to distinguish between different social processes, i.e. processes that result in one behavioural equilibrium and processes where many equilibria occur, acquiring a hallmark of culture and normative behaviour: arbitrariness in behaviour between groups. This allows us to talk about the wide space of behavioural regularities in a more granular fashion.

The definition also allows for the study of punishment/incentivisation on behaviour, whether it is genetically or culturally encoded. This allows us to also examine the effect of punishment on genetic evolution as well as on cultural evolution. This is an important aspect of the definition, as we believe how normative processes shape biological behaviour is understudied.

Finally, we argue that the permissive definitions (that fit under the minimal normative umbrella) allow us to appreciate the basic building blocks of normative cognition that may be excluded by more stringent definitions i.e. those requiring cognitive abilities, therefore allowing us to trace the evolution of normative cognition across phyla in a bottom-up fashion \cite{westra_pluralistic_2022,westra_search_2024,de_waal_towards_2010}. Following Westra et al, 2024 this normative framework is agnostic to the psychological mechanisms involved in normativity, therefore avoiding any bias that may mean we miss relevant phenomena because they lack human psychological capacities.

\subsection{Normative punishment and its impact on the evolution of mood}

In this section, we will give a definition of punishment and argue it applies to processes that occur across the animal kingdom. 

There could be several candidate mechanisms that maintain multiple equilibria normative regularities, but in this paper, we will investigate normative punishment. We define punishment as an action an agent takes that negatively affects another agent and is conditionally triggered by that same agent violating an explicit criterion, e.g. taking too much resource.
This negatively affects the agent's evolutionary success and, therefore, the propagation of their genetic or cultural strategy based on this criterion from the punishing agent. For example, the negative effect could be decreasing their chances of survival or ability to reproduce (so they do not propagate genetically), or their ability to spread their cultural profile, such as diminishing their status so they are not copied by others (thus limiting the spread of their behaviour culturally). 

Using the above definition of punishment, we can see that punishment does occur not only in higher-level organisms like humans, but across phyla \cite{raihani_punishment_2012}, even in simple organisms like insects \cite{wenseleers_enforced_2006} and bacteria \cite{huang_toxin-mediated_2023}. For example, e.g bacteria reduce the evolutionary success of free riders by releasing antibiotics, therefore killing and impeding free riders from passing on their behavioural phenotype to the next generation. We can now assert that punishment, according to our definition, is present in many species and is ergo, widely conserved in evolution (or occurred many times). Therefore, its influence must be considered in an evolutionary account of mood.

%We could in theory, choose from many forms of cognition and study how punishment effects it but we settle on mood for the following reasons:

In our experiments, we will verify if punishment (under our definition) results in multiple behavioural equilibria, therefore making it qualitatively different to other social processes, e.g. competition, and in our view, truly normative. We will then compare how mood evolves and emerges under punishment, i.e. explicit regulation of anothers' fitness and other non-multi-equilibria social processes, i.e. competition - implicit regulation of anothers' fitness and compare the effect of these processes on the evolution of the mood mechanisms.

%\clearpage

\section{Methods}

\subsection{Model overview}

We use an evolutionary multi-agent simulation with agents who have evolvable mood mechanisms that modulate their behaviour. Agents (N=100) take turns to consume a dynamically replenishing resource (eat step) and punish (sanction step). This is because this is a spaceless simulation with no topology, so agents punish a random subset of the agents by subtracting from their energy.
Energy level will determine procreation, and thus determine which agent parameters become more prevalent in the population. After initialisation, the simulation proceeds in rounds. Each round has a different, randomised order of all agents, and each of the following steps is performed in the following order for 2,000 time steps. Each condition is repeated 1000 times, and metrics are averaged over runs. And, at the start of each simulation, all agent values, including internal values and weights of the mood mechanism, are randomised.

\subsection{Eat} 
On their turn, each agent tries to consume an amount of resources of $\mu_{eat}$, which is determined by internal Bite Size ($B_{bite}$), modified by their current mood.
 This value is added to their internal energy and removed from the global resource level. If there are no resources left, the agent gets no energy.  If all agents eat at a higher $B_{bite}$, the environment will not be able to support as many agents; thereby, exhibiting 'tragedy of the commons' dynamics. A higher value is therefore considered more greedy or selfish \cite{hardin_tragedy_1968}. 
%\clearpage
\subsection{Sanction}
During their turn, the agent observes the actually consumed resources of the 10 previous agents. Each agent checks if the previous agents ate more than its own internal Sanctions Threshold ($B_{sanction}$) modified into $\mu_{sanction}$ based on their mood mechanism state (see below); if so, it sanctions them. In other words, $\mu_{sanction}$ is the amount of resource taken by another agent that an agent tolerates before punishing that agent, i.e. what an agent finds acceptable in terms of other agents acquiring resources. Sanctioning means the agent reduces the other agent's internal energy by its own Sanction Damage ($S_{damadge}$) which is set to 0.6 for all runs, and it pays a sanctioning cost of $0.1 \cdot S_{damadge}$, which is subtracted from its own energy level. 

\subsection{Metabolise}
Each agent has their energy level reduced by $0.1$ during each round.

\subsection{Death and Reproduction}
During the reproduction step, all agents with energy below 0 are removed. Then any agent with an energy level larger than 10 gets to reproduce. Reproduction means that we generate a copy of the agent with the same traits,  mutated with a $0.1$ chance (this is a global parameter that applies to all traits being mutated). Mutations are adding Gaussian noise with $0$ mean and $1$ variance to each of the agents' parameters. The energy of the child and parent are both set at half the parent's prior energy level. 
Further, to ensure evolution does not stop when populations are stable, we assume at each simulation step that each agent has a 1/100 chance of dying. Therefore, this means that each agent has, on average, a 100-time-step life span instead of being immortal, allowing for other agents to be born, and therefore allowing variation through mutation.

\subsection{Environment}

The environment comprises of a resource that replenishes dynamically with on average 10 units per time-step and is distributed across time in a sinusoidal manner with a peak of 20 units and a trough of 0 units.

\subsection{Initialisation}
At the beginning of the simulation, we create N = 100 agents and set the resource level to 1000 units. Each agent's internal values for $B_{bite}$ and $B_{sanction}$ are initialised to uniformly random values between $0.0$ and $1.0$. And the mood mechanism values are initialised between  $-1$.0 and $1.0$ for perceptual and behavioural weights and between $0.0$ and $1.0$ for $\alpha$ and $\beta$ parameters (see above). Each agent's energy level is set to 10.

\subsection{Model of mood: an affective layer to modulate behaviour}

We define \textit{mood} as the underlying experience of feeling, emotion or attachment. It is a catch-all terms a wide range of emotional/hormonal states and can be positive (e.g., happiness, joy, excitement) or negative (e.g., sadness, anger, fear, disgust). It is also less specific than an emotion: it captures this essence of how well things are going in the recent time window of the organism, and biases perception and action in an adaptive manner. Mood in our model is a fluctuating (positive or negative) affective state in our model \cite{russell_core_2003}. 

We use \textit{mood mechanism} to mean the tendency of an individual toward something, i.e. psychological inclination/disposition, and it can mean the environment’s power of disposing the individual to certain modes of interaction. Inspired by this, we define how mood is affected by stimuli (environment) and how mood, in turn, affects behaviour as a mood mechanism. We take inspiration for our model of mood from the empirically verified model of mood \cite{eldar_interaction_2015,eldar_mood_2016}.  More specifically, Eldar et al describe mood state as a scalar summary of "how well things are going": an average of recent reward error (positive or negative) over recent experiences in their case. We define it as average energy gain or loss in our case (it is our model’s equivalent of reward, see $S_{energy-gain}$ stimuli). We choose this model as it is a reification, capturing the essence of mood cognition instead of focusing on any particular emotion in any particular organism. Those core features being: stimuli (both internal and external) influencing a fluctuating mood state that in turn modulates behavioural and perceptual tendencies. We also add other stimuli as those chosen in \cite{eldar_mood_2016}. We chose these stimuli based on the sensory information available to the agents that we believe to be important regarding their survival in the simulated environment (both physical and social). For example, hunger, being punished, seeing others violating a norm, and seeing others die. This list is not exhaustive, and in theory you could add more (hence the dotted lines between $Stimulus_2$ and $Stimulus_n$). We randomly initialise the importance of these stimuli and the direction of their effect on mood as a way of seeing when evolution (if at all) utilises them. In other words, we let our evolutionary process specify the functional relationship between its senses and its mood and in turn, its mood and its behaviour. This functional specification is what we call a mood mechanism. This allows us to see different kinds of mood mechanisms emerging without presupposing a mechanism.  This allows us to explore how different types of social processes e.g. normative process like punishment pattern mood, produce different mood mechanisms.

% and add the ability for evolution to sculpt it: We don’t specify how stimuli affects mood and, consequently, behaviour and instead evolution specify the mood mechanism of the agents

Description of sensory stimuli:

\begin{itemize}
  \item $S_{hunger}$: How much of the agents bite size(how much you desire to eat) is unfulfilled $S_{hunger} = \mu_{eat} - resource $
  \item $S_{injured}$: how much energy loss was inflicted just from being damaged by sanctions.
  \item $S_{others-near-death}$: ratio of neighbours with less than 1 energy: (likely to die)
  \item $S_{others-near-birth}$: ratio of neighbours with more than 10 energy: (likely to give birth)
  \item $S_{others-being-punished}$: energy lost to punishment by the neighbours of the focal agent.
  \item $S_{energy-gain}$: The average energy gain/loss in the last 10 steps, this is similar to the model as proposed by \cite{eldar_mood_2016}.
  \item $S_{others-violations}$: the ratio of those around you violating the norm.
  \item $S_{dummy}$: This is an arbitrary fake stimulus that is not causally connected to the world in any way. We introduced this to analyse the dynamics of arbitrary traits and use it as a baseline to compare the other stimuli.

\end{itemize}

Now that we have described the overview of how the mood model operates, we will describe it in detail. What follows is a formal description of how mood affects behaviour in our model, with the weights (mood mechanism) affecting the direction and strength of this effect. In more detail, we describe how mood is updated based on stimuli and their evolvable weights, how mood updates with time and how mood consequently affects behaviour also through evolvable weights (see Figure 1 for overview). 

Equation (1) shows how different stimuli $S$ through weights $W$ affect how much the mood changes each time step ($\mathfrak{M}_{\delta}$), the $i$ is the index over the different sensors and their associated weights:

\begin{equation} \label{eu_eqn}
\mathfrak{M}_{\delta}  = \ \sum_{i=1}^{N} s \cdot W_n 
\end{equation}

In equation (2) the output of equation (1) $\mathfrak{M}_{\delta}$ (the change of mood)  is added to the current mood $\mathfrak{M}_{t}$, but is modified by $\alpha$ [0-1], another parameter that determines how much stimuli affect mood in general. The parameter $\beta$ determines how much mood decays each turn. This creates the mood of the next time step $\mathfrak{M}_{t+1}$.

\begin{equation} \label{eu_eqn}
\mathfrak{M}_{t+1} = (\mathfrak{M}_{t} + (\mathfrak{M}_{\delta} \cdot \alpha )) \cdot  \beta
\end{equation}

In equations (3) and (4) we see how mood $\mathfrak{M}$ affects internal behavioural values $B_{bite}$ (Bite size) and $B_{sanction}$ (Sanction Threshold), it does so with two weights (represented here as $w_B$ and $w_S$ behaviours to produce $\mu_{eat}$ and $\mu_{sanction}$.

\begin{equation}
\label{eu_eqn}
\mu_{eat} = B_{bite} + (\mathfrak{M_t} \cdot  w_B) 
\end{equation}
\begin{equation}
\label{eu_eqn}
\mu_{sanction} = B_{sanction} + (\mathfrak{M_t} \cdot  w_S) 
\end{equation}

\begin{figure*}[hbt!]

\[\begin{tikzcd}
	{S_n} &&&&&& \mu_{eat} \\
	{S_2} &&& Mood \\
	{S_1} &&&&&& \mu_{sanction}
	\arrow["{W_n}"{description}, from=1-1, to=2-4]
	\arrow[dotted, no head, from=2-1, to=1-1]
	\arrow["{W_2}"{description}, from=2-1, to=2-4]
	\arrow["{W_{bite}}"{description}, from=2-4, to=1-7]
	\arrow["{W_{sanction}}"{description}, from=2-4, to=3-7]
	\arrow["{W_1}"{description}, from=3-1, to=2-4]
\end{tikzcd}\]
\caption{An evolvable model of mood mechanism. Where different stimuli ($S$) and how they affect $Mood$ (fluctuating affective state), and how mood affects behaviour, with weights ($W$) affecting the direction and strength of these effects. Therefore, the weights define agents' mood mechanism. Note: the dotted lines between $S_2$ and $S_n$ imply that there are more stimuli that can be in-between  $S_2$ and $S_n$.}
\end{figure*}

\begin{table*}[hbt!]
\centering
\caption{Table of model parameters  
}
\resizebox{\columnwidth}{!}{
\begin{tabular}{@{}llll@{}}
\toprule
Parameter                       & Definition                                                                                     & Used value     & Justification                                                                                                                \\ \midrule
Sanction damage ($D$)           & The amount of energy a sanction damages the receiver of the punishment.                       & 0.6 (0.1-1 tested, see appendix)           & Chosen because it balances limited damage with norm compliance\\
Reproduction threshold          & The amount of energy an agent needs to have to reproduce.                                      & 10             & Chosen for population stability, i.e. when it is too low, population regulation is very difficult as agents reproduce too much. \\
Agent metabolism                & Amount of energy agents consume per time step, not taking into account punishment.             & 0.1            & Chosen so that the amount of resource in the environment is scarce enough to introduce a population regulation problem.\\
Mutation rate                   & The probability that each agent variable will "mutate" when reproduction occurs.               & 0.1            &                                                                                                                              \\
Mutation standard deviation     & Standard deviation used to mutate.                                                             & 1              &                                                                                                                              \\
Random death incidence rate     & Chance at each time step of an agent dying so that its life span is on average 100 time steps. & 0.01           & Chosen to ensure the population gets replenished and variation is introduced in the population.                              \\
Number of other agents punished & Number of agents the focal agent can see and punish.                                           & 10             &                                                                                                                              \\
Sanction cost to punisher       & The amount of energy consumed by the agent carrying out the punishment.                        & $0.1$ $\cdot$  $S_{damadge}$ &                                                                                                                             Reasonable that sanction also implies cost to sanctioner\\
Average growth rate             & Average replenishment of resource.& 10             & Chosen so that the amount of resource in the environment is scarce enough to introduce a population regulation problem.              \\
Resource at t = 0               & -                                                                                              & 1000           &                                                                                                                              \\
Number of agents at t = 0       & -                                                                                              & 100            & Chosen to ensure a sufficient amount of diversity among agents.                                                              \\
Energy of agents at t = 0       & -                                                                                              & 10             &                                                                                                                              \\
Period of environmental change  & -                                                                                              & 200 time steps & \\                                                                                                                          
\end{tabular}
}%
\end{table*}
%\clearpage
\subsection{Population level metrics}
\begin{itemize}
\item Hunger: the population average of how much of each agent's bite size (how much you desire to eat) is unfulfilled $Hunger := max(0, \mu_{eat} - resource)$. 
\item Injuries from sanctions: average energy loss across the population was just from being damaged by sanctions.

\end{itemize}

\subsection{Notes on evolution and model of mood}

Our model of evolution is not a complete model of evolution, but merely attempts to show how optimisation propagates the type strategies that climb the adaptive gradient.

%We are not interested in any particular affective phenomenon i.e. a specific emotion or hormonal mechanism. We model mood as an aggregate level phenomena and assume a particular reified model and not a specific one; many hormones play a role in mood but we are interested in the aggregate effect of these mechanisms here. Also, we take a model that has been empirically verified as a starting point \cite{eldar_interaction_2015,eldar_mood_2016}. 

%Further, we evolve multiple stimuli and let our evolution process specify each stimuli's role in behaviour instead of presupposing it. This allows us to examine how evolution couples mechanism to the various aspects of the social and environmental niche.

The model lends itself to both cultural and genetic interpretations, since more successful strategies (attaining more resources) become more common in the population. In the genetic interpretation, genes determining behaviour are passed down with mutation and are more likely to be passed down if they are successful at attaining a resource. In the cultural interpretation, it can be seen as strategies that are passed down culturally with a transmission error, with more successful strategies being imitated more to become the most common. The cultural interpretation allows for us to investigate the effect of normative punishment on norm emergence, and the genetic interpretation, which allows us to examine the effect of normative punishment on the evolution of mood. Further, this is a spaceless simulation with no topology. This was done in order to focus on the effects of normative punishment on the evolution of mood and not have it complicated by network topologies.
Further, we emphasise that the weights of the mood mechanism are being evolved, not the behaviours: $\mu_{eat} $ and $\mu_{sanction}$. They are produced by the evolved mood mechanism.
%\clearpage
\section{Results}

\subsection{Distinguishing between normative and non-normative social processes using the multiple equilibrium criterion}




In this section, we will reiterate our definition of an axiomatic normative regularity and operationalise it in our simulation. We will compare two conditions:
\begin{enumerate}
    \item Competition - implicit regulation of another's fitness.
    \item Punishment - explicit regulation of another's fitness.
\end{enumerate} 

We assess whether punishment does indeed produce multiple equilibria and therefore differentiate it from other social processes.

We define criteria for axiomatic normativity as \textit{patterns of behavioural conformity within a community} with the term "patterns of behavioural conformity", in our definition consisting of two criteria:

\begin{enumerate}

  \item The behaviour converges and stabilises: Do traits decrease in terms of their variance across the population from where they began, and do the average behaviours stabilise? We compute this by taking a mean average over the variance of a given trait in each randomly initiated run.

  \item The behaviour the population stabilises at is arbitrary across runs to a certain extent, i.e. multiple equilibria across runs are reached: This criterion ensures the resultant behaviour is not fully due to environmental or social scaffolding, meaning when the behaviour is the only rational/viable action given the environmental constraints \cite{westra_search_2024}. This would be indicated by repeated simulations stabilising at different average values i.e. there is a high variance of the means of each population.

\end{enumerate}

Now that we have articulated our definition, we will use it to interpret our simulation results, to confirm if explicit punishment processes and implicit ones lead to the dynamics predicted by the axiomatic normativity. 
In the explicit punishment condition: Agents can punish each other according to their mood modulated behaviour $\mu_{sanction}$, which determines what $\mu_{eat}$ from another agent triggers a punishment. In the competition condition: Agents only have $\mu_{eat}$ and socially interact indirectly through the shared resource.

\begin{figure*}[hbt!]
    \centering
    \subfigure[]{\includegraphics[width=1\textwidth]{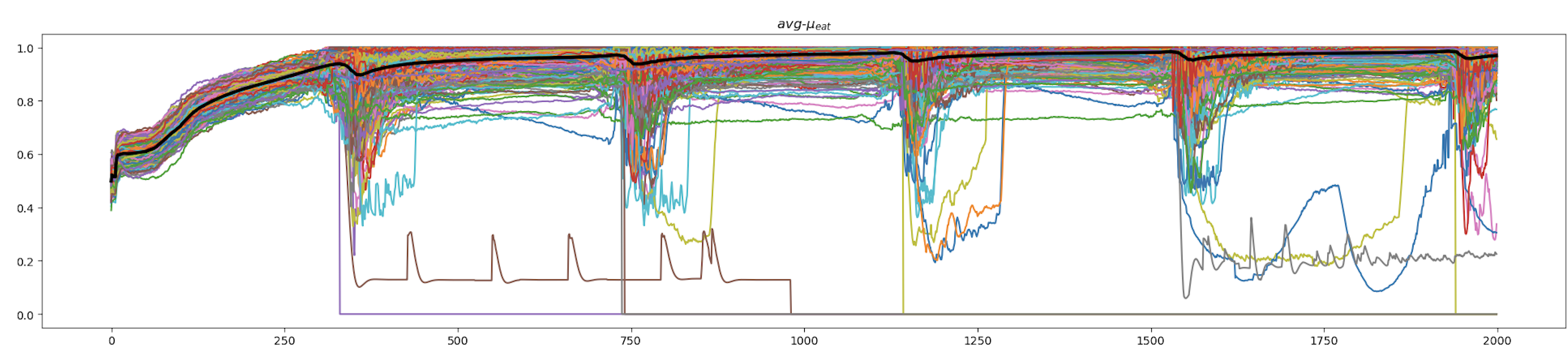}} 
    \subfigure[]{\includegraphics[width=1\textwidth]{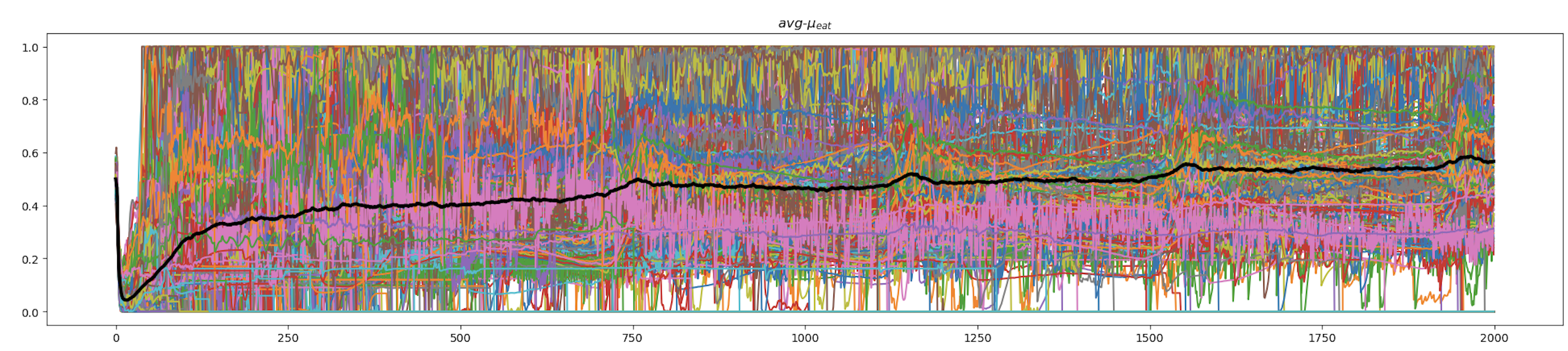}} 
    
    \caption{This figure depicts the average $\mu_{eat}$ for different simulation runs over time, with the black line being the average across runs and each coloured line the average of a population within a run. Comparing the competition and punishment conditions, we see that competition tends to one equilibrium (the maximum value), whereas the punishment condition results in a wide range of equilibria that cover the whole spectrum of $\mu_{eat}$.
    (a) Competition: In brief, in competition there is a positive correlation between avg-$W_{hunger}$ (bad mood) and avg-$W_{bite}$ indicating a  "bad mood → consume more" response    (b) Punishment (axiomatic normativity): In brief, there is a positive correlation between avg-$W_{energy-gain}$ (good mood) and avg-$W_{bite}$ indicating a "bad mood → consume less" mechanism. Additionally, there is a positive correlation avg-$W_{others-being-punished}$ vs. avg-$W_{bite}$, indicating that others will lower their consumption when they see others being punished.}
    \label{fig:eatmuavg}
\end{figure*}

Overall, we see in Table 2 that the punishment condition satisfies both conditions 1 and 2 of our definition. This is because we see a reduction in average variance of behaviour, signalling convergence, but still have variance between means, signalling arbitrariness.  Therefore, punishment results in behavioural regularities qualifying as axiomatically normative. Whereas implicit punishment i.e. competition, satisfies only condition 1. convergence not 2. arbitrariness, therefore not qualifying as an axiomatically normative process. We can conclude our definition works for distinguishing between these two processes in simulation (see also Figure\ref{fig:eatmuavg}).

%\begin{table*}[hbt!]
%\centering
%\caption{Variance of mean (extent of multi equilibria) and mean of variances (extent of behavioural convergence) for both the normative punishment and competition condition.}
%\label{tab:my-table}
%\begin{tabular}{|l|ll|ll|}
%\hline
%Traits                       & \multicolumn{2}{l|}{\begin{tabular}[c]{@{}l@{}}mean of variance:\\ extent to which a population\\ has converged on a behaviour\end{tabular}} & \multicolumn{2}{l|}{\begin{tabular}[c]{@{}l@{}}variance of mean:\\ extent of arbitrariness/multi equilibria \\ i.e. environmental/social scaffolding\end{tabular}} \\ \hline
%                             & \multicolumn{1}{l|}{competition }                                            & punishment& \multicolumn{1}{l|}{competition }                                               & punishment\\ \hline
%$\mu_{eat}$     & \multicolumn{1}{l|}{0.005}                                                  & 0.076                                                          & \multicolumn{1}{l|}{0.002}                                                     & 0.038                                                             \\ \hline
%$\mu_{sanction}$ & \multicolumn{1}{l|}{N/A}                                                    & 0.073                                                          & \multicolumn{1}{l|}{N/A}                                                       & 0.070                                                             \\ \hline
%\end{tabular}
%\end{table*}

\begin{table*}[hbt!]
\centering
\caption{Variance of mean (extent of multi equilibria) and mean of variances (extent of behavioural convergence) for both the normative punishment and competition condition after 2000 simulation steps. The start of simulation values are also displayed to provide a comparison. We see that for both competition and punishment conditions, the mean of the variance of $eat_{\mu}$ decreases, thereby satisfying criterion 1 of our normative regularity definition. However, when we look at the variance of the mean, we see that it is almost zero in the competition condition (as individuals all converge to high $eat_{\mu}$. However, in the punishment condition, we see much larger levels of variance across the means (an order of magnitude more), therefore maintaining a certain degree of arbitrariness. Therefore, only the punishment condition results in a behavioural regularity that meets both our criteria for being a normative regularity.}
\label{tab:my-table}
\begin{tabular}{|l|lll|ll|}
\hline
Traits                       & \multicolumn{3}{l|}{\begin{tabular}[c]{@{}l@{}}mean of variance:\\ extent to which a population\\ has converged on a behaviour\end{tabular}} & \multicolumn{2}{l|}{\begin{tabular}[c]{@{}l@{}}variance of mean:\\ extent of arbitrariness \\ i.e. environmental /social scaffolding\end{tabular}} \\ \hline
                             & \multicolumn{1}{c|}{\begin{tabular}[c]{@{}c@{}}start of\\ simulation\end{tabular}}     & \multicolumn{1}{l|}{competition}    & punishment    & \multicolumn{1}{c|}{competition}                                         & \multicolumn{1}{c|}{punishment}                                         \\ \hline
$eat_{\mu}$ (behaviours)     & \multicolumn{1}{l|}{0.29}                                                              & \multicolumn{1}{l|}{0.005}          & 0.076         & \multicolumn{1}{l|}{0.002}                                               & 0.038                                                                   \\ \hline
$sanction_{\mu}$ (behaviour) & \multicolumn{1}{l|}{0.29}                                                              & \multicolumn{1}{l|}{N/A}            & 0.074         & \multicolumn{1}{l|}{N/A}                                                 & 0.070                                                                   \\ \hline
\end{tabular}
\end{table*}

 Strong selection for $\mu_{eat}$ (how much resource the agent takes from the environment), which leads to low variances in the means in both competition and punishment conditions, Table \ref{tab:my-table}. However, there is more variance of the mean for $\mu_{eat}$ in the punishment condition. This suggests that this is not merely environmental scaffolding but a true normative regularity. Whereas the variance of mean for $\mu_{eat}$ in the non-minimally normative condition is almost 0 suggesting that even though there is a reduction in variance, it is mostly down to environmental and social scaffolding and not truly normative. However, one could argue that both these behaviours are socially maintained (as per \cite{westra_search_2024}); in the competition case, agents compete with each other for resources and therefore implicitly regulate each other's behaviour through leaving each other less resources in the environment. So you could argue that emergent behaviours constitute "normative regularities" based on the definition of \cite{westra_search_2024}. However, our definition allows distinction between regularities based on whether the system tends to multiple equilibria or not.
The key difference between punishment and competition is that the former allows for many equilibria, but the latter tends to one equilibrium: i.e. maximal $\mu_{eat}$, since that is the only direction of change that is permitted given the competition of the other agents, the system will tend to everyone being maximally selfish.
So even though competition indirectly influences others' behaviour through competition for resources in the environment, it's only through explicit punishment, when the agents are directly affecting each other's actions based on $\mu_{eat}$, that one gets many equilibria (as indicated by the higher variance between populations). This is due to the tension between punishment (explicit regulation of another's cultural or biological fitness) and the drive to maximise utility, where many possible outcomes exist, as there are many ways one can balance these two pressures i.e. one can choose to take the punishment and then eat a lot to counteract it or one can choose to eat a little and not get punished, and then there are many such combinations in-between that are viable. Thus, through normative punishment, there are now many possible equilibria that can be reached due to this push and pull, which is more like true social norms in that they are determined by the group and somewhat arbitrary \cite{anagnou_effect_2023,rai_social_2024} vs non-normative social processes like competition that tend to one equilibrium.

%ADD IN GRPAHS ON MEANS TO SHOW VARIANCE "bad mood → consume more" response, creating a tragedy of the commons that leads to resource depletion and population collapse. Under punishment, agents evolve a "bad mood → consume less" mechanism

\subsection{Different (social) mood mechanism emerges under punishment vs competition}

Under non-normative social process(competition) vs. normative process (punishment), we see two different versions of mood mechanism emerge, which rely on different stimuli to affect behaviour in different ways (Figure 3). The correlation matrix illustrates how the stimuli weights and behaviour weights are wired and therefore how a given stimulus will affect mood and subsequently how the mood will affect the behaviour.

We see that in both conditions, information about energy (being taken in, i.e. as signalled by hunger or energy-gain) is used. Thereby confirming, something not entirely surprising, that using the energy gained in the last state is useful for surviving.

We see in the competition case (non normative) that we have a strong positive correlation with the avg-$W_{hunger}$ in relation to the avg-$W_{bite}$ (Figure 5 c) and weak positive weights with respect to avg-$W_{energy-gain}$ vs avg-$W_{bite}$ and avg-$W_{hunger}$ vs. avg-$W_{energy-gain}$ (Figure 5 a,b). This means that agents, when they lose energy, tend to consume more. This losing energy state could be argued to be negative mood since losing energy tracks with being in an unfavourable state. \footnote{We have not pre-defined what a “good” mood is. Since both the input weights and output weights can evolve to values from -1.0 to 1.0, the system has one functional symmetry, i.e. we could invert all input and all output weights and would get the same functionality. As a result, there is no ''a priori'' definition of what a ''good'' mood is. For later interventions, we consider the good mood direction to be the one that correlates with positive energy gain as an input. The reason we do this is that the individuals who managed to survive in the simulation all had mood mechanisms that made them behave in a way to gain energy rather than lose it; otherwise they would not have survived. This also fits well with mood definitions that generally define it as “how well an organism is doing” \cite{eldar_mood_2016}, but again, we have let this arise through evolution and not hand-code it ourselves.} The mechanism we get from the population with competition is in line with what we would expect from a homeostatic model of affect: in response to negative mood, exert a large response to shift the system back into a favourable state (to not be hungry in our case). This is similar to states like hunger or stress in the literature that use negative feedback like stress hormone cortisol \cite{schulkin_rethinking_2002, khan_long-term_2022}.

We see that this is not the case with the punishment condition (normative case), where we have no correlation for avg-$W_{bite}$ vs. avg-$W_{hunger}$ and for avg-$W_{energy-gain}$ vs. avg-$W_{hunger}$ (Figure 4, b, c). Instead, we have a strong positive correlation for avg-$W_{energy-gain}$ vs avg-$W_{bite}$ (Figure 4, a), meaning agents eat more when they are gaining energy (arguably like a good mood), and they eat less when losing energy (arguably a bad mood). This behaviour seems irrational as agents will not increase their consumption when they are losing energy and are, therefore, in a more precarious state. However, if we look at this behaviour in the context of punishment, losing energy is also correlated with being punished. Therefore, in the context of punishment, it is rational for the system to lower its eating amount to avoid punishment and to increase it in the absence of punishment. This is further strengthened by the fact that there is a strong negative correlation between the avg-$W_{others-being-punished}$ vs. avg-$W_{bite}$, meaning agents, when they see others being punished, may indicate to them that they may be punished as well, so in response, they reduce the amount they are eating to avoid punishment.  
Another observation worth mentioning is the clustering of values away from 0 in a and d, and to a lesser extent e and f. It seems unlikely for a population to not arrive at a relation between these weights, suggesting that there is an adaptive advantage to having this relation vs. not having it.

This behaviour, eating more when an agent is in a good mood and less when the agent is in a bad mood, is the emergent encoding of a social preference, which takes into account the social consequences of taking too much resource. This social preference is encoded in how the mood mechanism works itself, with punishment rewiring what it means to be in a "good" or "bad" mood. We emphasise that this emerges through evolution and punishment alone, without us hand-coding the effects of stimuli on mood and mood on behaviour. Moreover, eating more in a favourable state (good mood) is similar to positive feedback-based hormonal mechanisms, e.g. social hormones like oxytocin, which are involved in social and non-social functions \cite{taylor_tend_2006,khan_long-term_2022}. Furthermore, the agents withdrawing from taking from the shared resource when in a "bad" mood, in order to avoid punishment, can be interpreted as a depressive mood.
According to the social risk hypothesis, depression represents an adaptive response to the perceived threat of exclusion from important social relationships that, over the course of evolution, have been critical to maintaining an individual's fitness prospects \cite{allen_darwinian_2006,allen_social_2003}.

Further, there is a negative correlation between  avg-$W_{energy-gain}$ vs. avg-$W_{sanction}$. This means that agents seem to punish less when they are losing energy, and they are more tolerant to deviance when they are in a "bad mood" (they are losing energy either because there is no food/they are punished) and punish more when they are losing energy (bad mood). This seems irrational, but again, if we look at this behaviour in the context of punishment, it allows for agents to minimise the punishment expenditure on energy when they are in a bad mood (they are losing energy either because there is no food/they are being punished) and therefore more efficient in a social environment \cite{hertwig_fast_2009}.

\begin{figure*}[hbt!]
    \centering
    \subfigure[a]{\includegraphics[width=0.42\textwidth]{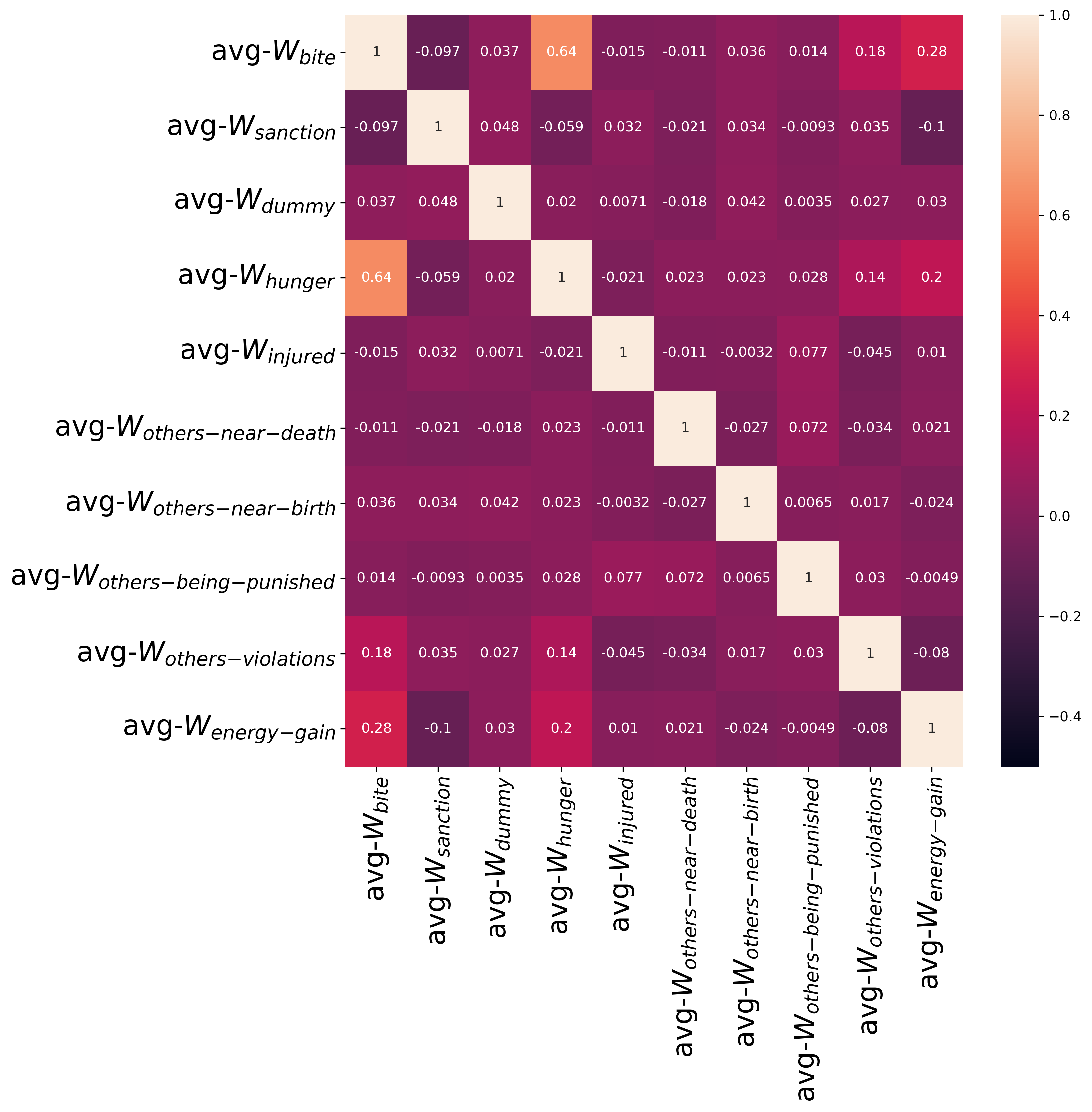}} 
    \subfigure[b]{\includegraphics[width=0.42\textwidth]{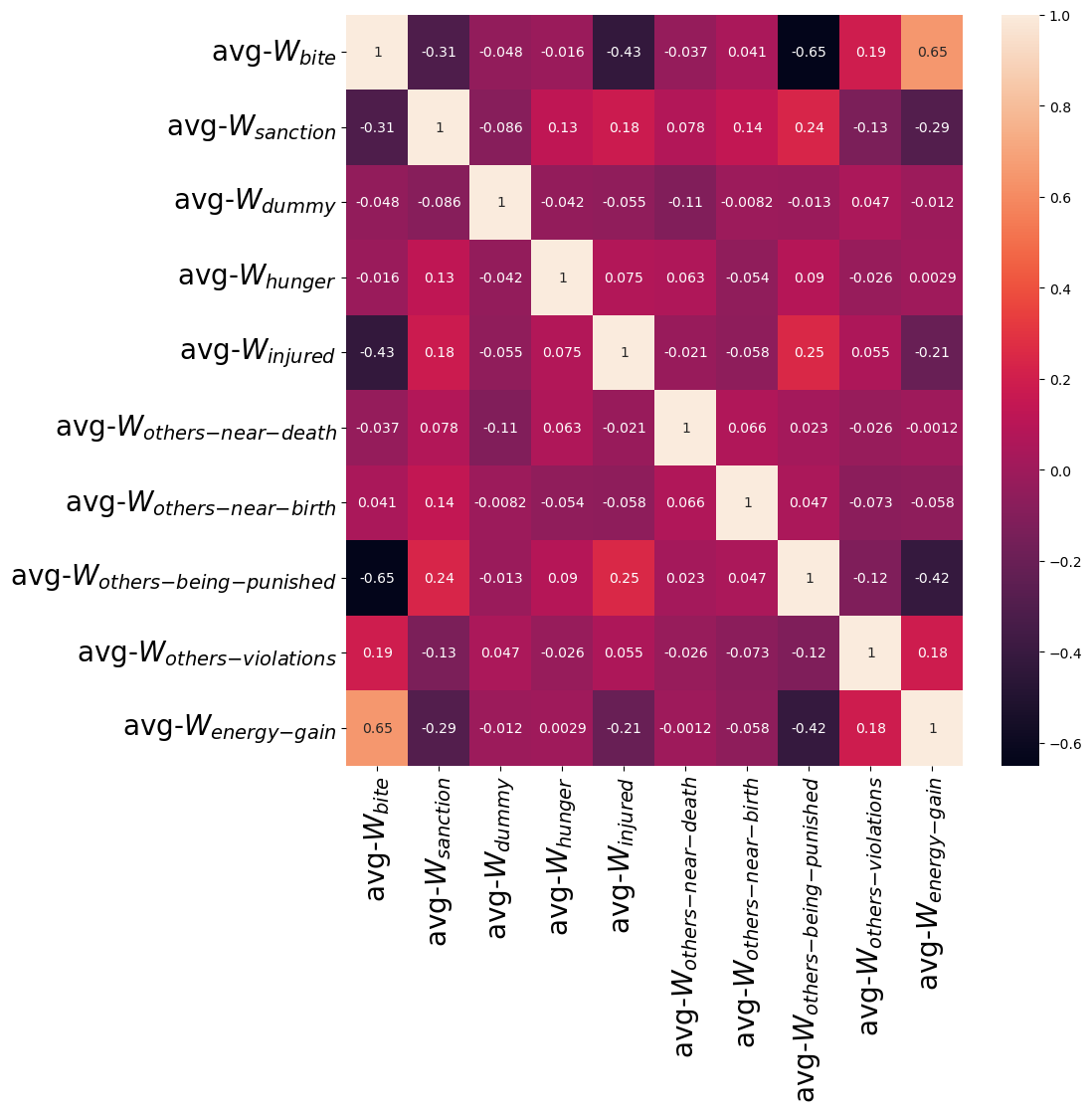}} 
    
    \caption{Here we see the effect of evolution on the correlations between different stimuli weights, behavioural weights which encode mood mechanism, i.e. how stimuli affect mood and how mood affects behaviour respectively (as well as other evolvable agent traits) and how they are correlated with one another across populations after evolution.
    (a) competition (b) punishment (axiomatic normativity)}
    \label{fig:}
\end{figure*}

\begin{figure*}[hbt!]
    \centering
    \subfigure[]{\includegraphics[width=0.24\textwidth]{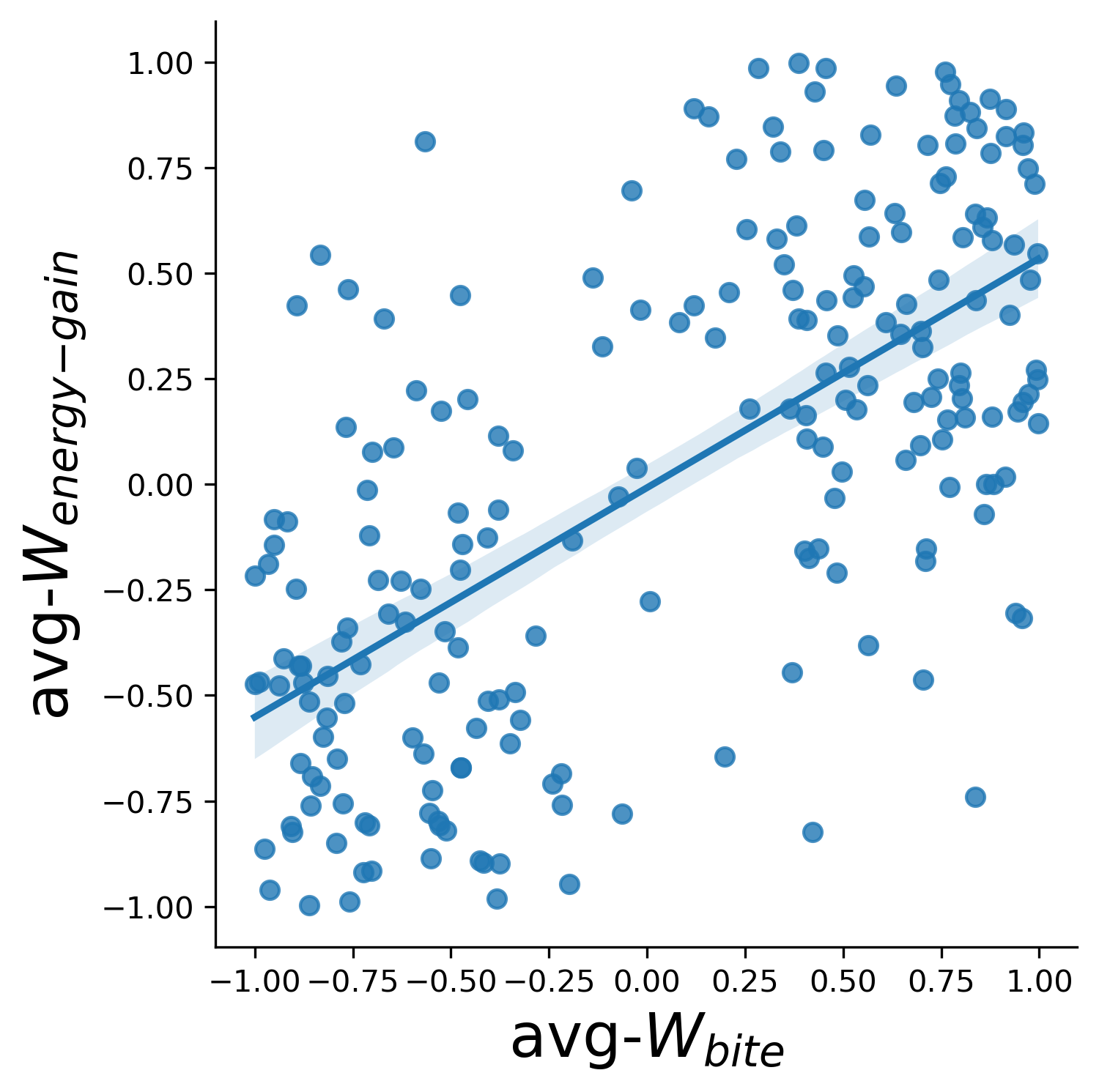}} 
    \subfigure[]{\includegraphics[width=0.24\textwidth]{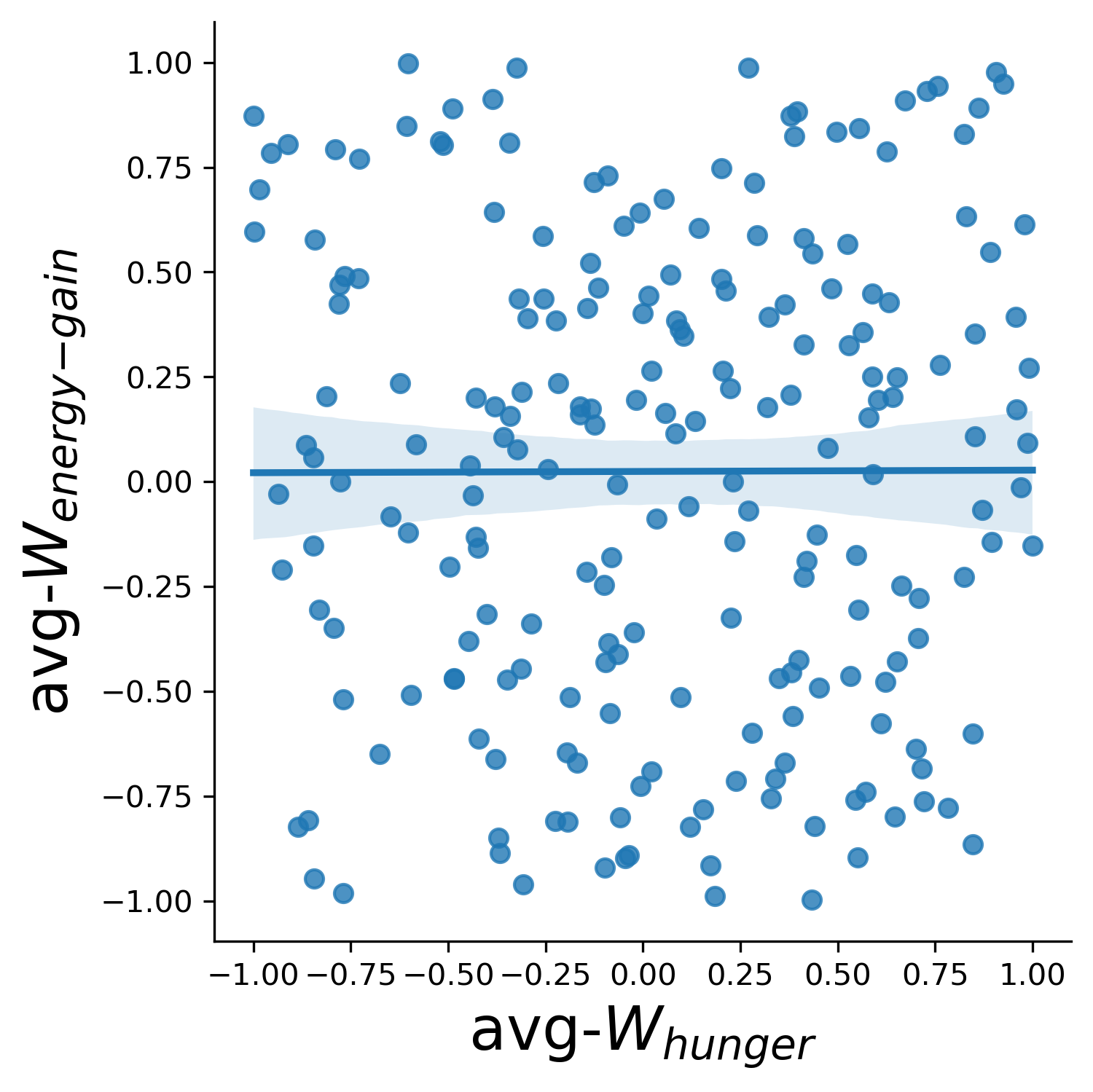}} 
    \subfigure[]{\includegraphics[width=0.24\textwidth]{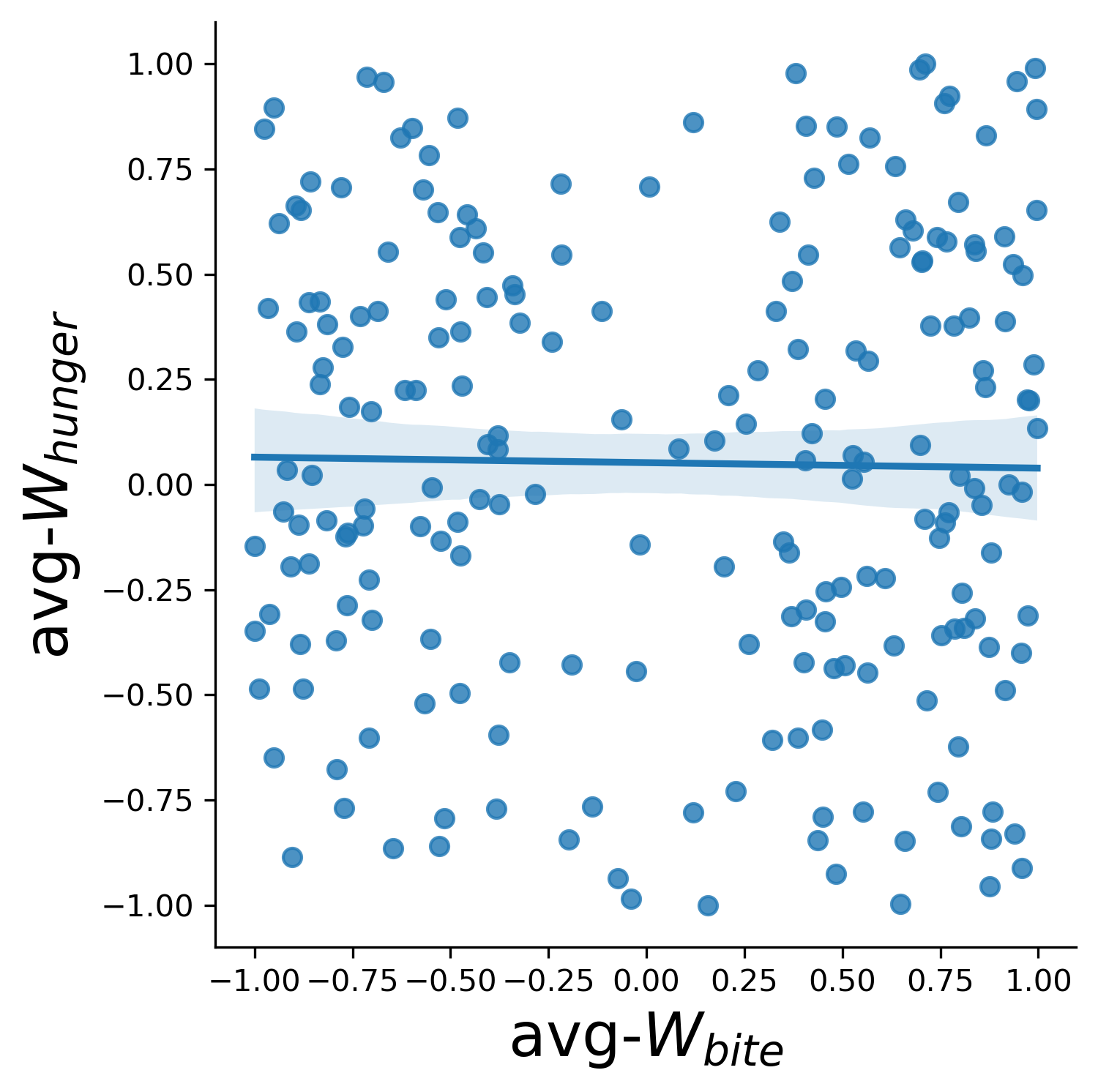}}
    
    \subfigure[]{\includegraphics[width=0.24\textwidth]{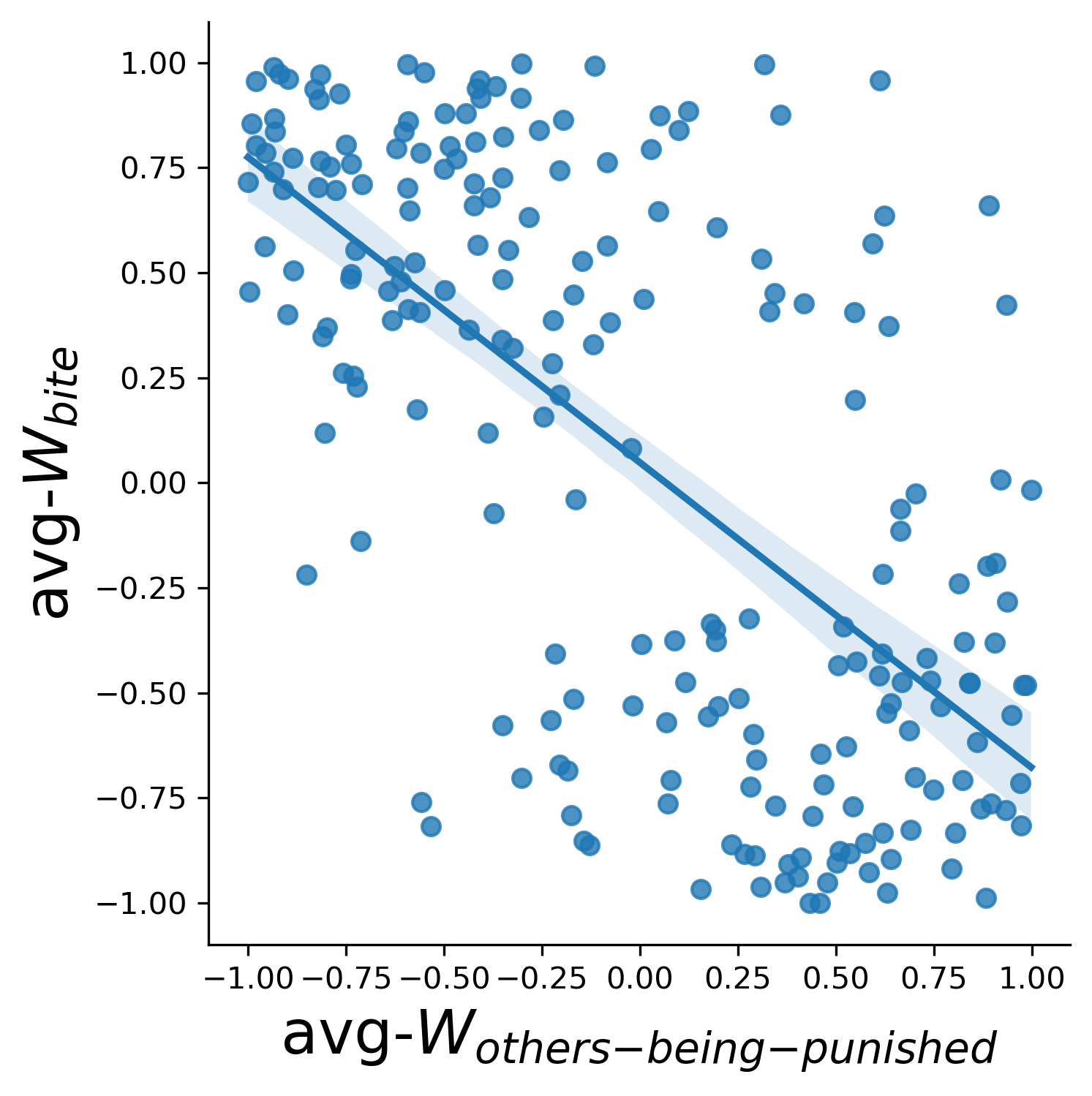}}
    \subfigure[]{\includegraphics[width=0.24\textwidth]{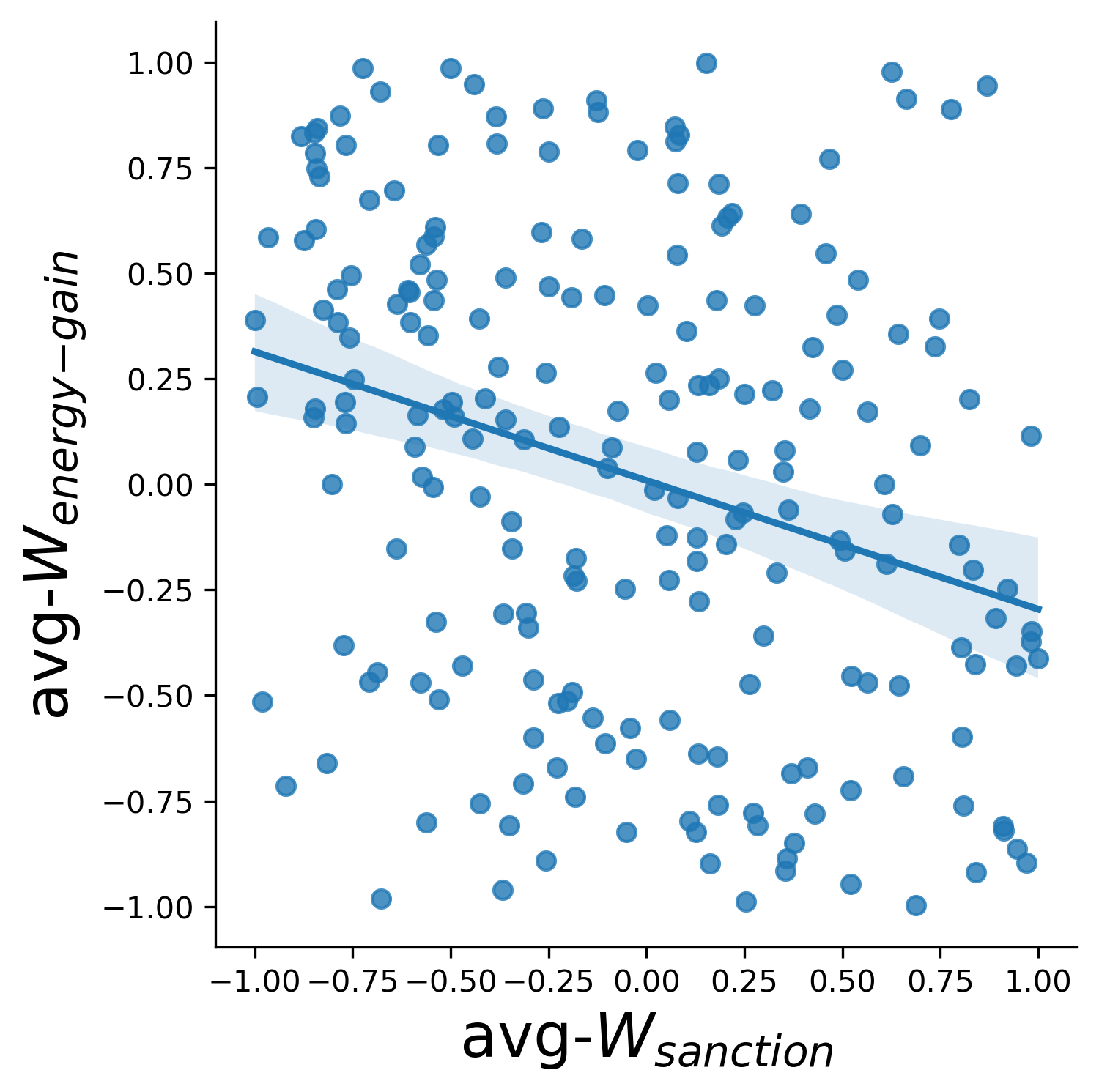}}
    \subfigure[]{\includegraphics[width=0.24\textwidth]{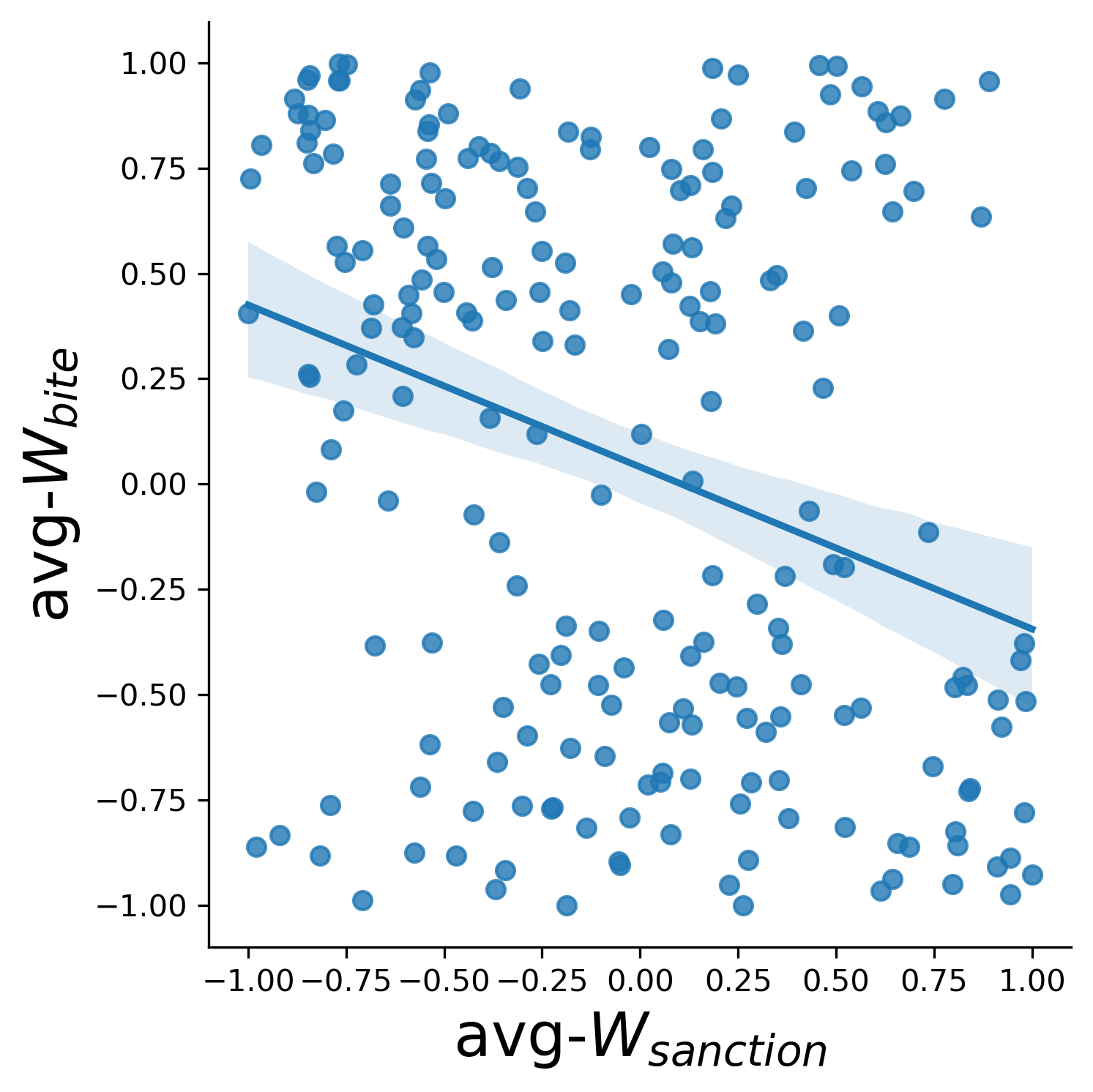}}
    \caption{This figure depicts various averages of different traits plotted against each other for the condition with punishment (normative) to see the relation between traits over different runs. Each blue circle is a different simulation run, and the blue line indicates the general trend. Here we see that avg-$W_{energy-gain}$ vs avg-$W_{bite}$ is the strongest positive correlation, indicating on average that agents will eat more if they are gaining energy and eat less if they are losing energy. Further, we see that there's a negative correlation between avg-$W_{sanction}$ vs avg-$W_{energy-gain}$, indicating that agents will punish more (since the sanction threshold is lower) when they are gaining energy.  We also see negative correlation for avg-$W_{others-being-punished}$ vs avg-$W_{bite}$, indicating agents will eat less if they see others being punished.
    (a) avg-$W_{energy-gain}$ vs avg-$W_{bite}$ (b) avg-$W_{hunger}$ vs avg-$W_{energy-gain}$ (c) avg-$W_{hunger}$ vs avg-$W_{bite}$ (d) avg-$W_{others-being-punished}$ vs avg-$W_{bite}$ (e) avg-$W_{sanction}$ vs avg-$W_{energy-gain}$ (f) avg-$W_{sanction}$ vs avg-$W_{bite}$}
    \label{fig:scatter_with_main}
\end{figure*}

\begin{figure*}[hbt!]
    \centering
    \subfigure[]{\includegraphics[width=0.24\textwidth]{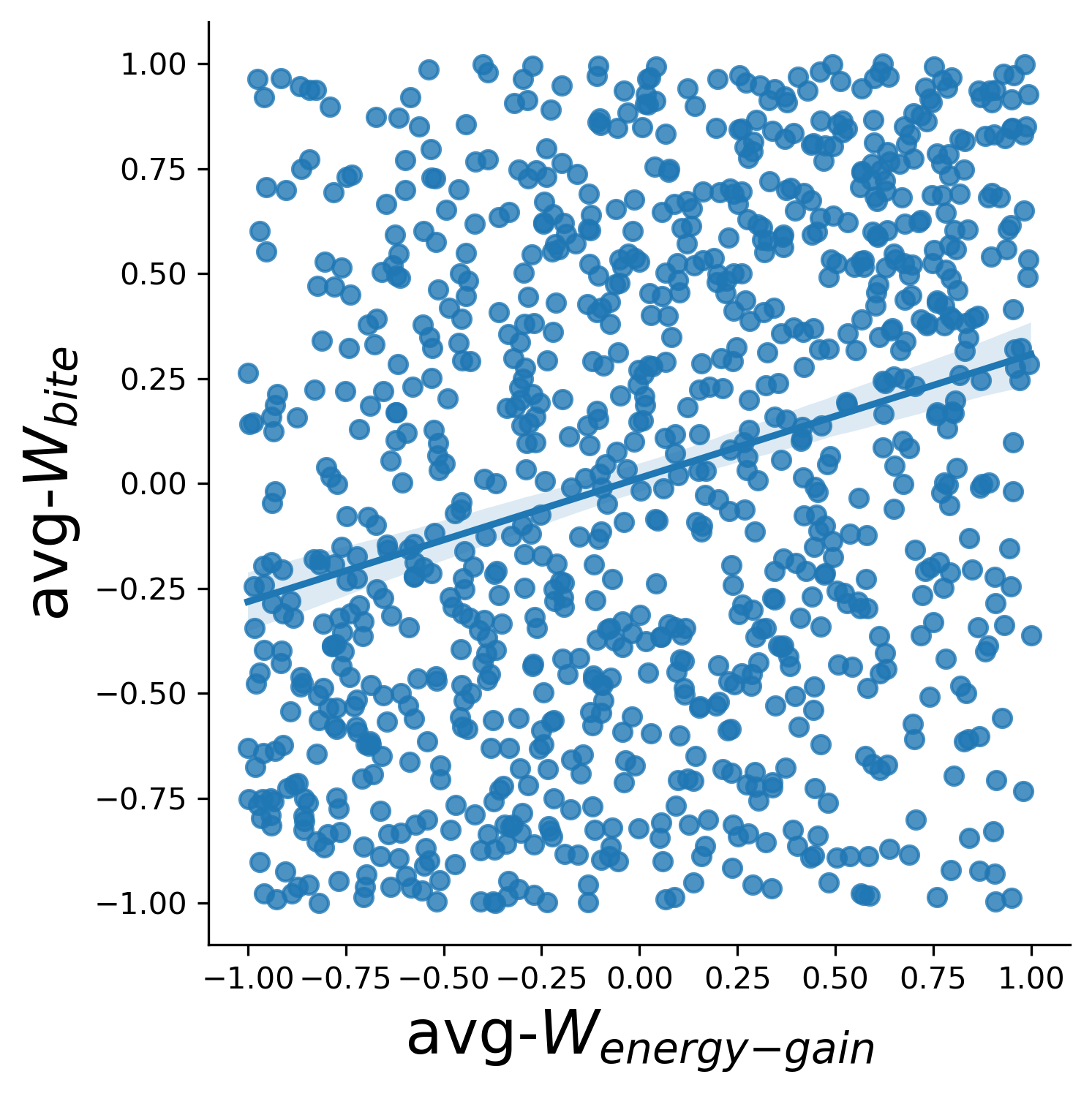}} 
    \subfigure[]{\includegraphics[width=0.24\textwidth]{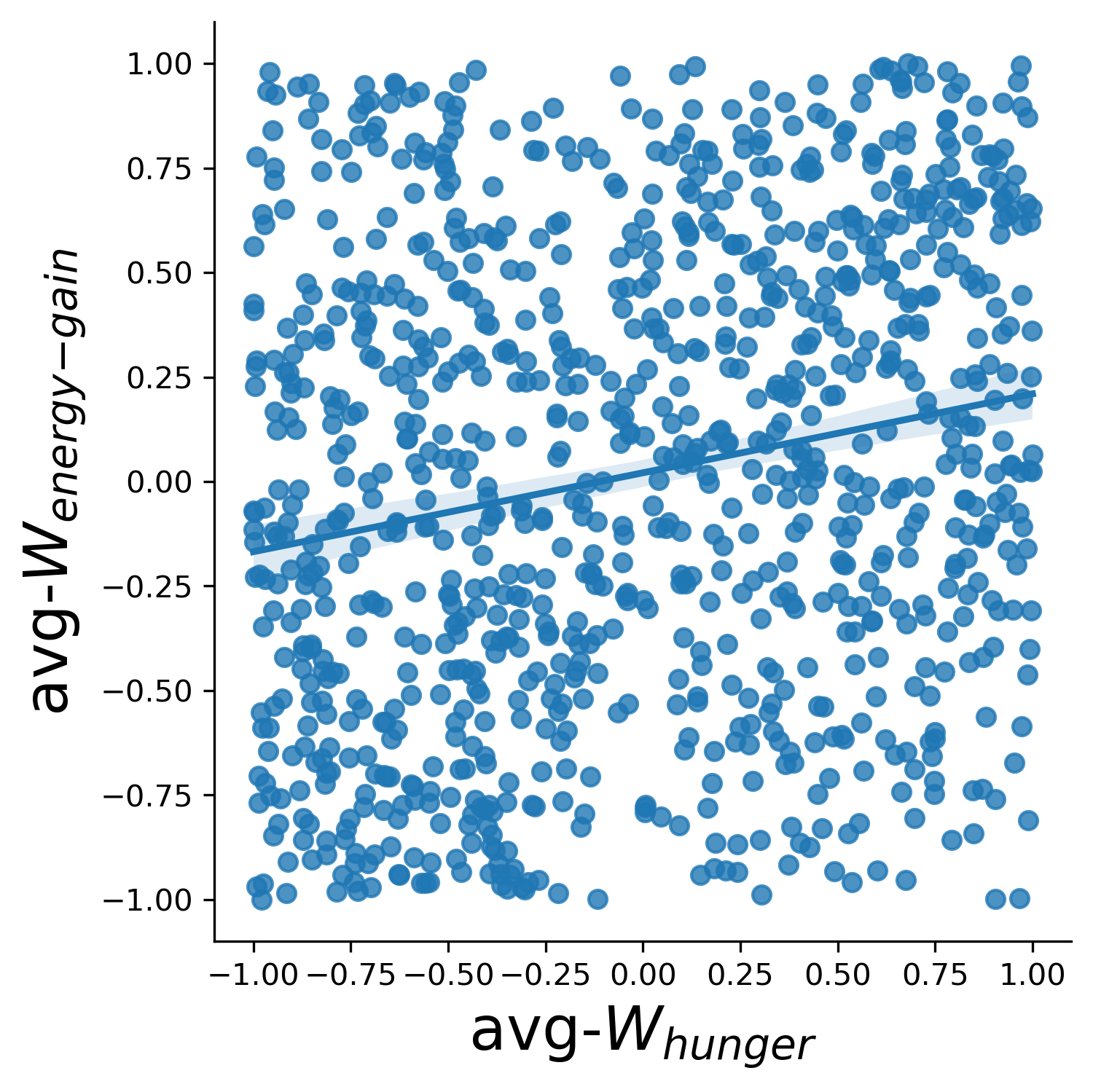}} 
    \subfigure[]{\includegraphics[width=0.24\textwidth]{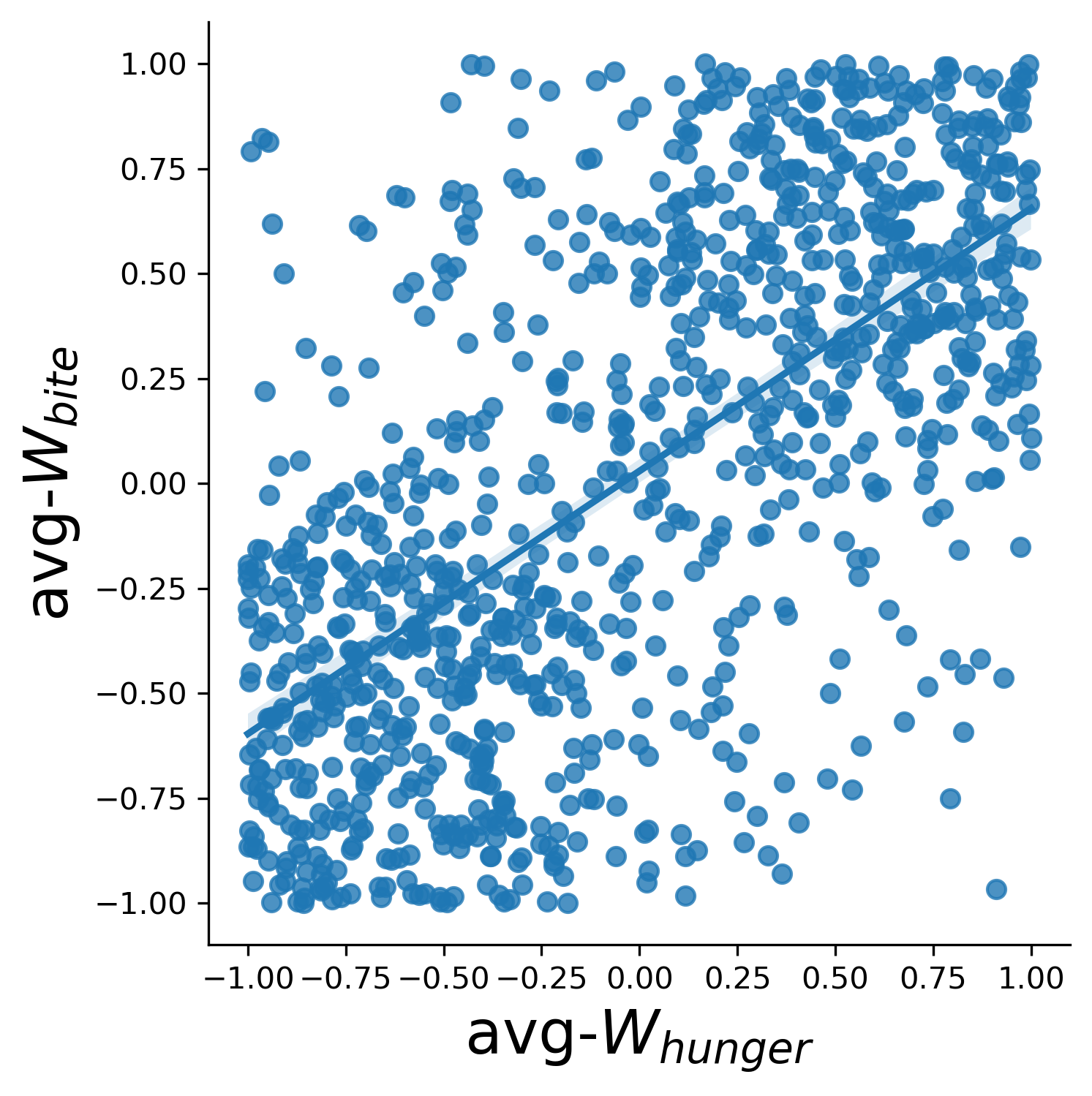}}
    \caption{This figure depicts various averages of different traits plotted against each other for the condition with competition (non-normative) to see the relation between traits over different runs. Each blue circle is a different simulation run, and the blue line indicates the general trend. Here we see, in contrast to the punishment condition, a strong correlation between avg-$W_{hunger}$ vs avg-$W_{bite}$ and weak correlations for avg-$W_{energy-gain}$ vs avg-$W_{bite}$ and avg-$W_{hunger}$ vs avg-$W_{energy-gain}$. This means that on average agents will eat more if they are in a hungry state and losing energy  (a) avg-$W_{energy-gain}$ vs avg-$W_{bite}$(b) avg-$W_{hunger}$ vs avg-$W_{energy-gain}$ (c) avg-$W_{hunger}$ vs avg-$W_{bite}$}
    \label{fig:scatter_WO_main}
\end{figure*}

\subsection{Mood mechanism that emerges with punishment avoids population collapse}
In the previous section, we showed that punishment patterned the mood mechanism, resulting in a behaviour that is socially rational - i.e. eat less when losing energy and eat more when gaining energy. This means the agents avoid eliciting punishment from each other.
Now we will examine what this does on the group level to the population dynamics and ascertain whether this is indeed beneficial to the group and therefore prosocial.

\begin{figure*}[hbt!]
    \centering
    \subfigure[a]{\includegraphics[width=0.75\textwidth]{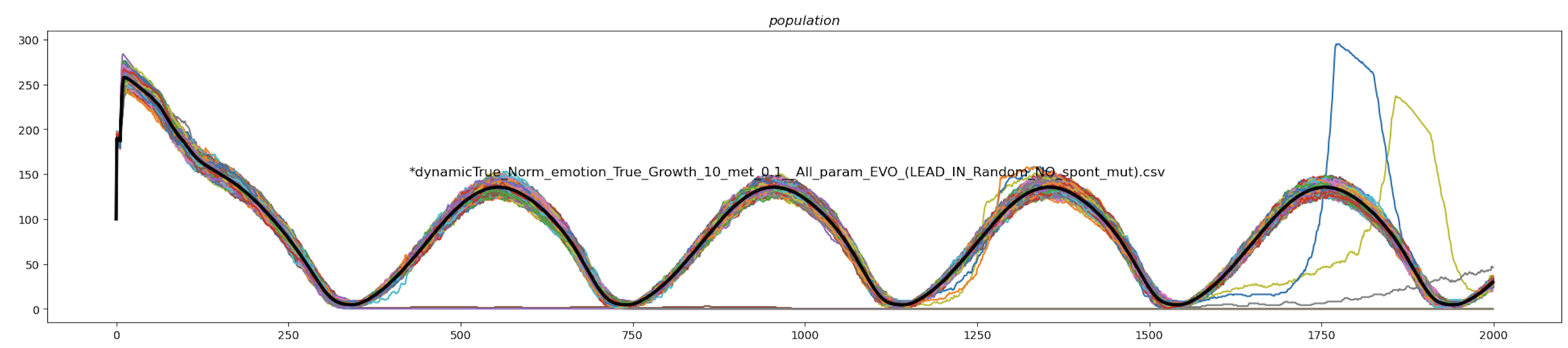}} 
    \subfigure[b]{\includegraphics[width=0.75\textwidth]{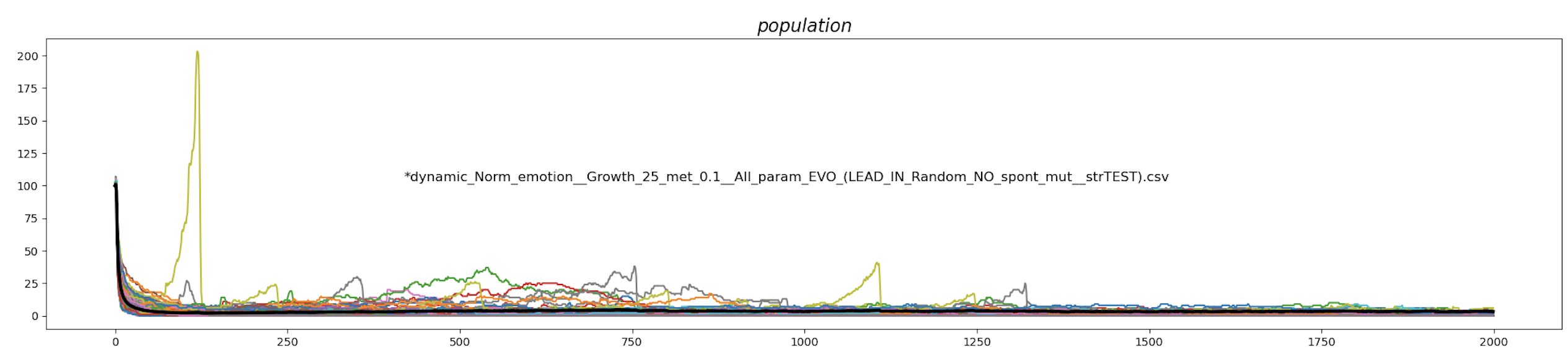}} 
    \subfigure[c]{\includegraphics[width=0.75\textwidth]{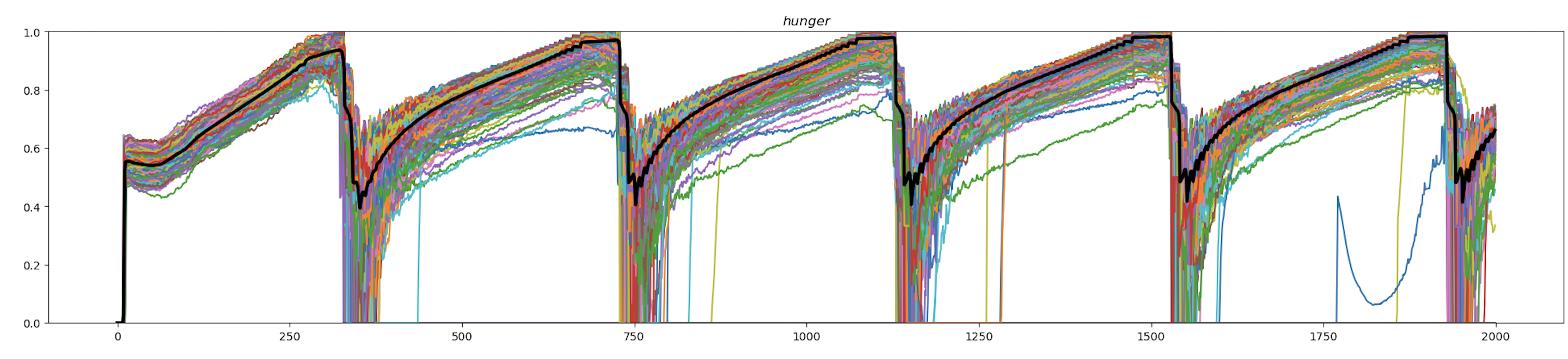}}
    \subfigure[d]{\includegraphics[width=0.75\textwidth]{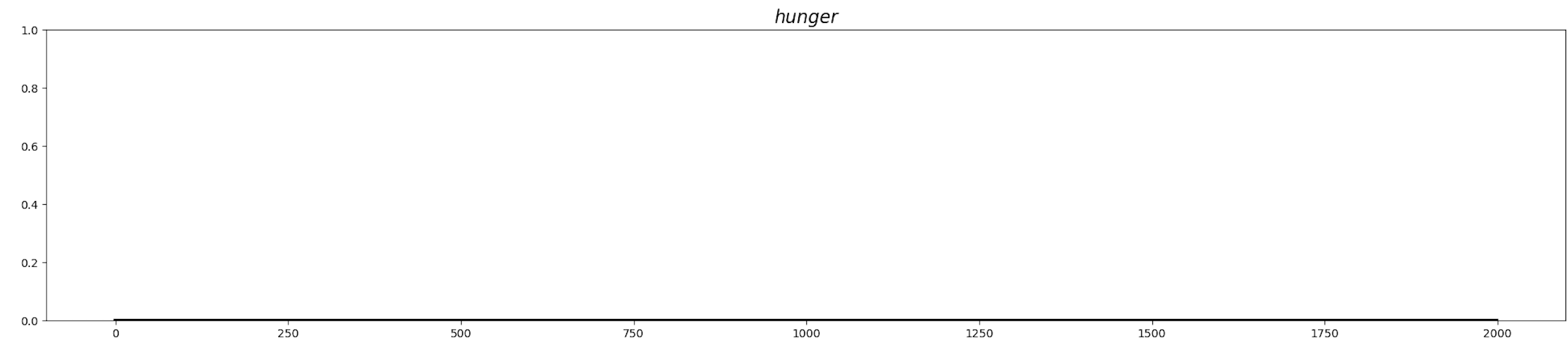}}
    \subfigure[e]{\includegraphics[width=0.75\textwidth]{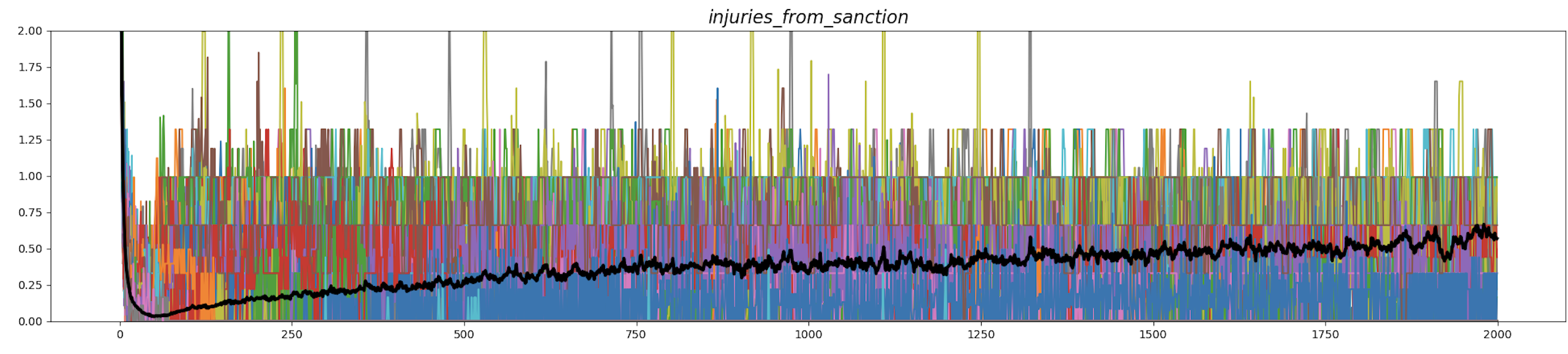}}
    \caption{This figure depicts the average population level traits for different simulation runs over time. Comparing the competition and punishment conditions. N.B. In the punishment condition, we plotted only runs that evolved high punishment and conserved energy compared to the competition condition see SI for full runs).
    a) Population of runs in competition condition b) Population of runs in punishment condition 
    c) Hunger of runs in competition condition d) Hunger of runs in punishment condition.
    The coloured line depicts the average of trait for individual runs and black line is average over all runs
    e) Injuries from sanction in punishment condition: this is the average energy lost to punishment across the population.}
    \label{fig:pop_level_metrics}
\end{figure*}

In the population plots for the competition condition (Figure \ref{fig:pop_level_metrics}a), we see that populations grow when there is plenty of resource; however, during periods of low resource the populations collapse due to overexploitation of the resource during the growth phase. 

In contrast, for a subset of runs in the punishment condition, agent populations do not overgrow during periods of high resource and therefore do not experience the same collapse of population due to overexploitation, Figure \ref{fig:pop_level_metrics}b. This is due to agents punishing each other, taking energy away from each other, and therefore not allowing the overgrowth that results in collapse of the population later on. We argue that this is a prosocial outcome from the perspective that overgrowth and then collapse result in many agent deaths.
This being said, an issue that emerges:  Since punishment is so costly to the population(Figure \ref{fig:pop_level_metrics}e), it results in an extreme inefficiency that means the populations remain very small.

\subsection{Mood injection to causally confirm effect of mood on behaviour}

Although the correlation matrix analysis sheds some light on how behaviour is organised, sometimes correlations can be spurious. So we decided to causally probe the simulation. 

\begin{figure*}[hbt!]
    \centering
    \includegraphics[width=1\textwidth]{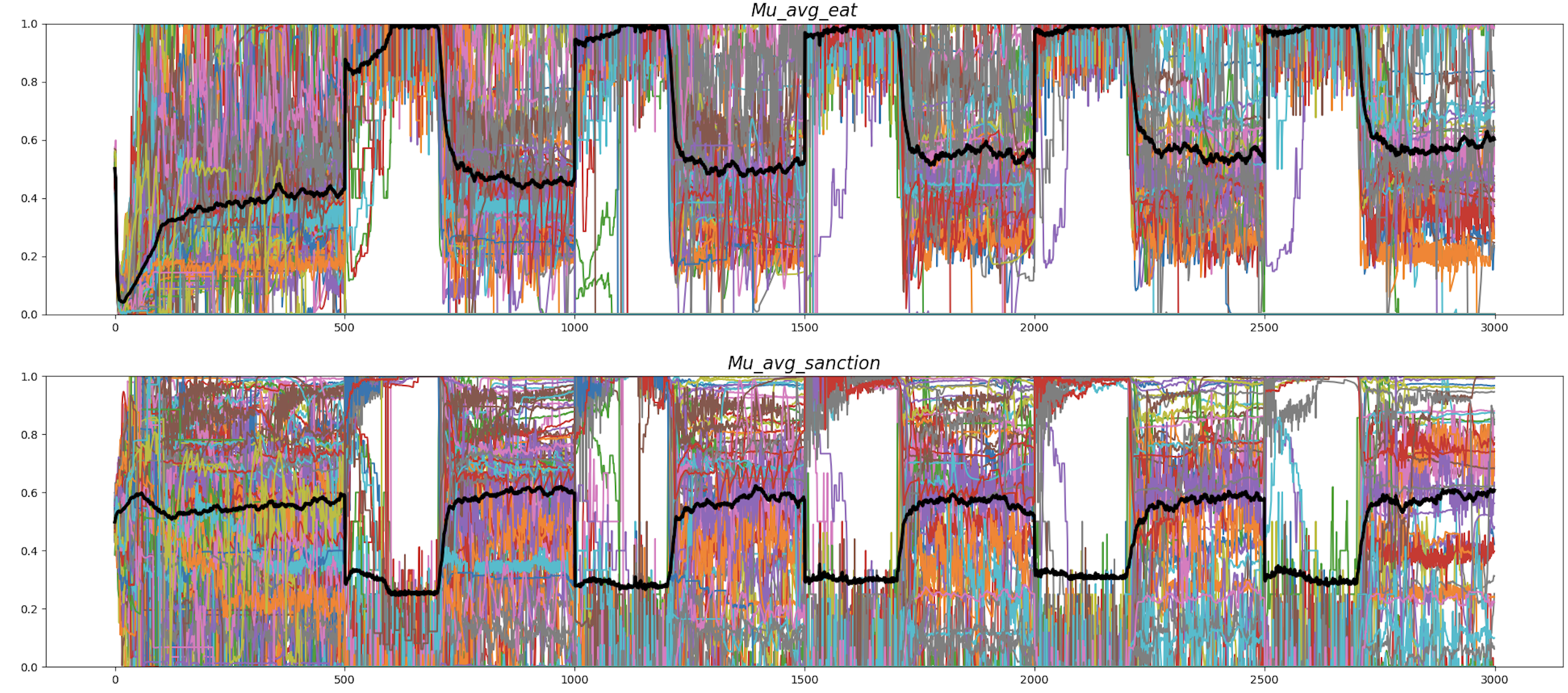}
    \caption{Mood injection: the average value for each population is plotted for two traits: $\mu_{eat}$ above and $\mu_{sanction}$ below. The average across the population is plotted as a black line. Injections of positive mood occur every 500 time steps for 200 steps. Since positive mood (feeling good) is defined by whether the weights for avg-$W_{energy-gain}$(average energy gain) are positive or negative, we injected negative mood if it was negative (to create a net positive) and positive if it was positive (again for net positive). This means that for any case, we are effectively injecting  positive mood into the system/} 
    \label{fig:DISPO_inject}
\end{figure*}

To do this, we artificially "injected" positive mood into the agents to see if there was a net effect on the behaviour. By this, we mean artificially making the mood of every agent positive to see how agents change their behaviour when they are in a positive mood.
We did this every 500 time steps, "injecting" 200 units of positive mood for 200 time steps into each agent, and these graphs show how this affected behaviour Figure \ref{fig:DISPO_inject}.
We confirm that agents in the punishment condition increase their $\mu_{eat}$ when they are in a good mood and decrease it when they are in a low mood, and that they punish more in a good mood and punish less in a bad mood. We see that, especially in the $\mu_{sanction}$ graph, not all populations respond in the same way when a positive mood is injected, e.g. some decrease their amount of punishment. However, we can see that on average, injecting mood mechanism has the effect described above (eat more and punish more in a good mood, eat less and punish less in a bad mood), Figure \ref{fig:DISPO_inject}. Therefore, we can conclude that punishment does indeed cause our model of mood to become patterned as seen here.
This population-level connection between mood and behaviour opens up the possibility of the agents influencing each other with mood signals. This is because agents in the population now have a similar behavioural response to one another when they are in a similar mood. For example, if an agent could emit a signal that made others feel "bad" by lowering their mood, they would get them to consume less resource and vice versa. This would be a more sophisticated method to regulate other agents' behaviour than physically punishing them. Signalling of this type is a more sophisticated way that organisms regulate each other's behaviour, and avoids the inefficiencies and detrimental aspects of physical punishment \cite{andrighetto_punish_2013}.
Therefore, punishment (axiomatic normativity), by patterning mood mechanism on the population level, has set the stage for the evolution of higher-order normative signalling. 

%%Minor results(maybe something for supplementary info ]

%qualifier that including this mechanism in the first place is a good idea
%-better survival than no emotion with social maintenance (see Anagnou et al, 2023).due to flexibility in a dynamic environment. with mood mechanism: 260 vs Without: 188 

%%-aligning mood mechanisms, seems in times of stress mood mechanisms become more aligned in SM /vs not aligned in NSM. In terms of subjective well being, this suggests that is more "fair" when it comes to mood. This is true even when variances are similar

\newpage
\section{Discussion}
In this article asked how our moods are entangled with social norms, i.e. how did social preferences emerge in our mood mechanisms evolutionarily? We argue that since norm-like processes (minimal normativity) are replete across phyla, it could be a possible selection pressure that leads to social preferences evolving in mood mechanisms; despite this, many accounts of mood remain individualist.

We decided to address this with an agent-based model in which agents have a mood mechanism whose vital parameters could be evolved, therefore allowing us to study the effect of different aspects of the physical and social environment (specifically, axiomatically normative punishment) on the evolution of mood mechanisms in an artificial setting.

\subsection{Summary of results}
%CHANGE THIS UP 
%1. We show that what emerges in the two different contexts (with social maintenance and without social maintenance) is different, which is the first result, taking into account social maintenance gives us a different picture of affect emergence in evolution.

%2.We propose a refinement of Westra and Andrews definition of social maintenance that distinguishes between indirect social maintenance e.g. competition between individuals which usually tends to one equilibria i.e. Nash equilibrium/race to the bottom and a richer direct social maintenance (explicit regulation of another's behaviour based on some criteria) that allows for many equilibria. Further we also distinguish between simple single variable norms e.g. walking speed and mechanistic norms, multiple variables that specify a behavioural mechanism, converging at the population level e.g. a culturally patterned mood mechanism.

\begin{itemize}

\item 1a. According to accounts of minimal normativity, we see normative processes across phyla.

\item 1b. We verify that punishment (explicit regulation of another's cultural or genetic phenotypic fitness) creates multiple equilibria and is therefore qualitatively different from other forms of social maintenance/social behaviour, e.g. competition, and therefore fundamentally normative.

\item 2. Use our axiomatic normativity definition to model the effect of normative punishment on the evolution of mood. We show that this could explain the emergence of social preference, i.e. how mood systems become patterned in such a way as to be altruistic (not just the binary uptake of a variable or not, as it's usually modelled).

\item 3. Show that minimally normative processes imprint themselves on the mood mechanism and allow mood mechanisms to become normative themselves, setting the stage for the evolution of mood signalling mechanisms. 
\end{itemize}
\subsection{Discussion of results}

We show that considering normative punishment in the evolution of mood has a strong effect, producing a qualitatively different mood mechanism compared to considering just competition. We distinguish between competition and punishment, showing that the latter is axiomatically normative. We define an axiomatically normative process as a process that gets agents to converge on a behaviour while producing multiple behavioural equilibria instead of just one. Thus, capturing a key aspect of normativity, a certain degree of arbitrariness. The fact that normative punishment is widely conserved across phyla \cite{raihani_punishment_2012}, even in simple organisms like insects \cite{wenseleers_enforced_2006} and bacteria \cite{huang_toxin-mediated_2023}, should urge us to consider the effects of axiomatically normative processes on the evolution, even in ancient mechanisms such as mood and other cognitive processes.

Using an evolvable model of mood, we show that evolution and punishment alone can result in the encoding of a social preference in a mood mechanism. This mood mechanism takes into account the negative social consequences of taking up too many resources, encoded in a distributed manner across the weights of the mood mechanism, and gets the agent to act in a socially adaptive manner. This is in contrast to other models of social preference, where preferences are often hardcoded into models (e.g normative concern/ guilt \cite{alexander_staller_introducing_2001, cimpeanu_evolutionary_2025} and explanations about how they emerge evolutionary from purely material concerns i.e. purely from individual level self interest are thin on the ground and assume trivial mappings between genes and behaviour that omit mechanism \cite{bowles_cooperative_2013,gavrilets_collective_2017}.
We provide a possible mechanism for how social motivations/preferences can evolve from purely material incentives and normative social interaction. In other words, social preferences are co-constructed in a decentralised manner through agents manipulating each other's fitness through normative punishment. This also goes to show that there is no objectively good mood mechanism, but that it depends on the interactions between the agent and the environment (in this case social environment). In other words, there is no "right way" to feel; it is subjective, it depends on what is adaptive given the context.

Further, our study shows how a social preference is "smeared" across the mechanism's weights and opens the door to talk about how norms may emerge in a mechanism rather than a particular behaviour specified by one value (e.g. selfishness). Thinking of norms as a collection of values that specify a socially patterned version of a psychological mechanism, in this case, mood. This then defines certain behavioural tendencies in certain situations and changes over time. This opens up considering how even very simple systems become socially patterned and offers a richer conception of norms that takes into account the underlying mechanism producing the behaviour and not only the behaviour itself. This is similar to the approach taken by the Evolutionary Game Theory community in the context of the genetic evolution of behaviour, studying the evolutionary dynamics mechanisms variables in an equation, that specify some sort of mechanism (e.g. Rescorla-Wagner learning) rather than a single variable that defines a strategy \cite{mcnamara_integrating_2009,taborsky_towards_2021,mcnamara_game_2022}.

The mood mechanism allows multiple concerns to be encoded in one, i.e. mood reflecting both lack of food but also the level of punishment. One could start thinking of mood as funnelling certain goals into one. One may feel bad when they do not get what they are individually interested in, but also when socially shamed (i.e. ones mood depends on what others think of you). If an agent maximises for feeling "good", then they are implicitly capturing multiple goals in one "utility" to maximise. This can be the basis for brains, or in this case, simple mood mechanisms becoming socially entangled and therefore socially rational \cite{hertwig_fast_2009}. Furthermore, agents withdrawing from taking from the shared resource when in a "bad" mood, in order to avoid punishment, can be interpreted as a depressive mood. According to the social risk hypothesis, depression represents an adaptive response to the perceived threat of exclusion from important social relationships that, over the course of evolution, have been critical to maintaining an individual's fitness prospects \cite{allen_darwinian_2006,allen_social_2003}.

We show through causal probing of our simulation that invoking negative or positive mood in the evolved agents results in agents reacting in a predictable manner: eating less when they are in a bad mood and more when they are in a positive mood ". This opens up signalling to each other to change their behaviour with the threat of punishment \cite{andrighetto_punish_2013}, i.e. direct effect on mood through negative signals (yelling at someone). This would mean that you could get the desired population regulation effect without the physical damage from the sanctions, Figure \ref{fig:DISPO_inject}.
This is possible because including punishment in the first place has resulted in social entanglement. That is, alignment of the way mood mechanisms work, meaning agents are socially entangled and have a common emotional response. This means that they can, in turn, use emotional (positive and negative) signals to influence each other. In effect, punishment has imprinted a kind of "normative logic" into the mood mechanism that allows more sophisticated mind shaping to take place without the need for agents to physically punish each other, which is costly.  This kind of mind shaping is commonly thought to occur in organisms only with complex cognitive faculties like Theory of Mind \cite{frith_theory_2005,zawidzki_mindshaping_2013}, but we suggest a plausible path where it can emerge from the simple ingredients of mood and normative punishment.

%c) We extend a known definition of norm emergence ,aka normative regularities \cite{westra_search_2024}, and provide an operationalisation of this in a model in a continuous norm context:
%1. The behaviour converges and stabilises, and 2. the behaviour the population stabilises at is arbitrary across runs to a certain extent, which enable us to distinguish cases of norm emergence through social maintenance and environmental scaffolding and provide a further refinement by distinguishing between indirect and direct social maintenance. Indirect social maintenance leads to one equilibrium (similar to environmental scaffolding) and the richer direct social maintenance results in many possible equilibria in behaviour. This difference is due to differing number of motivations affecting behaviour. In the indirect case, there is only one motivation (maximising resource intake) that governs an agents success. Therefore, in this case, strategies will tend to the maximum value as they compete against each other for survival. In the direct case, there are two competing motivations that an agent must balance, maximising resource intake but also not being punished, this results in a continuum of viable strategies that balance these motivations. This results in direct social maintenance leading to many equilibria, and we believe it captures an important aspect of normative behaviour in that it bears a certain degree of arbitrariness. 

\subsection{Advantage of our model vs other models}
Our model does not use discrete motivations as assumed in other models, i.e. \cite{alexander_staller_introducing_2001, khan_long-term_2022}, but allows them to be encoded in a distributed manner across a network. On the other hand, different from traditional neural network model,s which, while more abstract and with fewer assumptions, are harder to relate to the real world due to their opaqueness. Our model can relate back to real motivations through the use of affective language while still seeing how social preferences are “smeared” across a network instead of presupposing discrete motivations. Therefore, our model is a nice middle ground between these approaches.

\subsection{Limitations}

We have only included punishment (negative explicit regulation of others' fitness in our model. But, we also know that positive regulation may exist through incensation of good behaviour, for example, whose inclusion may allow for richer dynamics to emerge. 
Moreover, we have a single-dimensional view of mood, and there exist different models that have at least two dimensions other than valence and magnitude \cite{nettle_evolutionary_2012, trimmer_evolution_2013}. In the future, we can expand our model to see the effects of minimal normativity on these models of mood.

\subsection{Future work}

We showed that punishment and mood in the preserve of evolution resulted in alignment of mood mechanisms in a population. Meaning agents are socially entangled and have a common emotional response, which allows for higher-order mindshaping \cite{zawidzki_mindshaping_2013} through signalling \cite{andrighetto_punish_2013}. That is, direct effect on mood through negative signals (yelling at someone) to regulate one another. Future work could use simulations to see if it is possible to evolve mood signalling. This would help answer whether indeed this sophisticated mind shaping behaviour can emerge from the simple ingredients of normativity and mood. Possible complications of having evolved mood signalling would be interesting to consider. For example, it opens up the possibility of emotional free-riders, i.e. agents that manage to "unevolve" their response to the punishment signal to regulate themselves and start acting selfishly again. Or you may even get others abusing the signal for their own gain, reducing others' greed in order to reap the benefits, a sort of mood manipulation. This may impede mood signalling evolving at all, with agents evolving to have completely unpredictable responses in order not to be manipulated. For evolution to overcome this roadblock, we may need to posit, e.g. honest signalling mechanisms or commitment devices for something like this to evolve \cite{mercier_not_2022,kelly_yuck_2011}.

\printbibliography

%%%%%%%%%%%%%%%%%%%%%%%%%%%%%%%%%%%%%%%%%%%%%%
\section{Appendix}

% \usepackage{multirow} 
\begin{table*}[hbt]
\caption{Parameter exploration of sanction damage ($S_{damadge}$) and its effect on population level traits.}
\resizebox{\columnwidth}{!}{

\begin{tabular}{|l|lllllllllll|}
\hline
\multirow{2}{*}{Agent population level traits}                 & \multicolumn{11}{c|}{Sanction Damage ($S_{damadge}$)}                                                                                                                                                                                                                                                                        \\ \cline{2-12} 
                                                               & \multicolumn{1}{l|}{0}     & \multicolumn{1}{l|}{0.1}    & \multicolumn{1}{l|}{0.2}    & \multicolumn{1}{l|}{0.3}    & \multicolumn{1}{l|}{0.4}    & \multicolumn{1}{l|}{0.5}    & \multicolumn{1}{l|}{0.6}    & \multicolumn{1}{l|}{0.7}    & \multicolumn{1}{l|}{0.8}    & \multicolumn{1}{l|}{0.9}    & 1      \\ \hline
Agent population surival rate /1000 intial populations         & \multicolumn{1}{l|}{1000}  & \multicolumn{1}{l|}{782}    & \multicolumn{1}{l|}{688}    & \multicolumn{1}{l|}{475}    & \multicolumn{1}{l|}{340}    & \multicolumn{1}{l|}{258}    & \multicolumn{1}{l|}{187}    & \multicolumn{1}{l|}{181}    & \multicolumn{1}{l|}{152}    & \multicolumn{1}{l|}{155}    & 125    \\ \hline
avg-$W_{energy−gain}$ vs avg-$W_{bite}$      & \multicolumn{1}{l|}{0.28}  & \multicolumn{1}{l|}{0.09}   & \multicolumn{1}{l|}{0.43}   & \multicolumn{1}{l|}{0.66}   & \multicolumn{1}{l|}{0.73}   & \multicolumn{1}{l|}{0.7}    & \multicolumn{1}{l|}{0.65}   & \multicolumn{1}{l|}{0.68}   & \multicolumn{1}{l|}{0.62}   & \multicolumn{1}{l|}{0.72}   & 0.57   \\ \hline
avg-$W_{sanction}$ vs avg-$W_{energy−gain}$ & \multicolumn{1}{l|}{N/A}   & \multicolumn{1}{l|}{0.05}   & \multicolumn{1}{l|}{-0.14}  & \multicolumn{1}{l|}{-0.31}  & \multicolumn{1}{l|}{-0.38}  & \multicolumn{1}{l|}{-0.37}  & \multicolumn{1}{l|}{-0.29}  & \multicolumn{1}{l|}{-0.38}  & \multicolumn{1}{l|}{-0.1}   & \multicolumn{1}{l|}{-0.34}  & -0.19  \\ \hline
avg-$W_{hunger}$ vs avg-$W_{energy−gain}$       & \multicolumn{1}{l|}{0.64}  & \multicolumn{1}{l|}{0.13}   & \multicolumn{1}{l|}{0.07}   & \multicolumn{1}{l|}{0.06}   & \multicolumn{1}{l|}{0.04}   & \multicolumn{1}{l|}{0.03}   & \multicolumn{1}{l|}{-0.016} & \multicolumn{1}{l|}{0.11}   & \multicolumn{1}{l|}{0.07}   & \multicolumn{1}{l|}{0.072}  & -0.2   \\ \hline
$µ_{eat}$ mean of variance                                         & \multicolumn{1}{l|}{0.005} & \multicolumn{1}{l|}{0.0521} & \multicolumn{1}{l|}{0.0592} & \multicolumn{1}{l|}{0.072}  & \multicolumn{1}{l|}{0.0697} & \multicolumn{1}{l|}{0.0817} & \multicolumn{1}{l|}{0.076}  & \multicolumn{1}{l|}{0.0892} & \multicolumn{1}{l|}{0.0718} & \multicolumn{1}{l|}{0.0677} & 0.0674 \\ \hline
$µ_{eat}$ variance of mean                                         & \multicolumn{1}{l|}{0.002} & \multicolumn{1}{l|}{0.0269} & \multicolumn{1}{l|}{0.0391} & \multicolumn{1}{l|}{0.0473} & \multicolumn{1}{l|}{0.0468} & \multicolumn{1}{l|}{0.0436} & \multicolumn{1}{l|}{0.038}  & \multicolumn{1}{l|}{0.0255} & \multicolumn{1}{l|}{0.0267} & \multicolumn{1}{l|}{0.0292} & 0.0254 \\ \hline 
\end{tabular} %
}
\end{table*}

Here we describe a parameter exploration for sanction damage ($S_{damadge}$) (the amount of energy being taken away from an agent when they are punished) (Table 3). The first trend is that as sanction damage ($S_{damadge}$) increases, the population decreases. This is straightforwardly explained by the loss of energy due to larger $S_{damadge}$ hindering the growth of populations.
Secondly, traits for the population regulation mechanism peak at 0.4 when (avg-$W_{energy-gain}$ vs avg-$W_{bite}$ is positive and avg-$W_{sanction}$ vs avg-$W_{energy-gain}$ is negative), with the presence of the mechanism tapering off at high and low values of $D$. 
Lastly, traits for norm emergence also peak at 0.4 (norm emergence defined as low mean of variance and high variance of mean) with lower levels of norm emergence for low $S_{damadge}$ and high $S_{damadge}$.
This breaks down into two factors of norm emergence: First, the variance of the mean which peaks at 0.4 and tapers of at high and low levels of $D$. When $S_{damadge}$ is low, we expect norm emergence to be less pronounced since the motivation for agents to conform is weaker. However, what is unexpected is that the variance of mean also decreases at high levels of punishment, this may be due to a smaller number of populations at higher $S_{damadge}$ which decreases the possible variance between them. 
Second, the mean of variance, which is lowest (strong norm emergence) at low and high levels of $S_{damadge}$ with it being higher (weaker norm emergence). Having a lower variance of the mean makes sense when punishment is harsher, but this does not explain why it appears to be lowest when there is no punishment. The low variance in behaviour is a consequence of indirect social maintenance, in this case all the agents are competing against each other and therefore tending to maximum bite size, thus explaining the low variance is behaviour as all agents tend to one equilibrium. Since there is no variance in the behaviour across populations, it does not count as norm emergence.

Overall, there seems to be a trade-off between stronger norm emergence and the presence of population regulation mechanism and the survival rate of populations, of which 0.4 is the best (judging best is not straightforward, depending on what observer values. In this case strong norm emergence without too low a survival rate). However, even in this better case, the argument raised in the main body of the paper on still stands: the cost of the population regulation mechanism is that the survival rate is lower. Therefore it would be advantageous for agent populations to evolve a way of affecting each other’s behaviour through affective signals instead of directly through physical damage.

\end{document}